%% file: sample-manuscript.tex
\documentclass[acmsmall]{acmart} 

\AtBeginDocument{%
  \providecommand\BibTeX{{%
    \normalfont B\kern-0.5em{\scshape i\kern-0.25em b}\kern-0.8em\TeX}}}

\input{Commands_Acronyms.tex}

\begin{document}
\sisetup{tight-spacing=true}

\title[Hybrid Modular Redundancy]{Hybrid Modular Redundancy: Exploring Modular Redundancy Approaches in RISC-V Multi-Core Computing Clusters for Reliable Processing in Space}

\author{Michael Rogenmoser}
\authornote{Both authors contributed equally to this research.}
\email{michaero@iis.ee.ethz.ch}
\orcid{0000-0003-4622-4862}
\affiliation{%
  \institution{ETH Z\"{u}rich}
  \city{Z\"{u}rich}
  \country{Switzerland}
}

\author{Yvan Tortorella}
\authornotemark[1]
\orcid{0000-0001-8248-5731}
\email{yvan.tortorella@unibo.it}
\affiliation{%
  \institution{Universit\`{a} di Bologna}
  \city{Bologna}
  \country{Italy}
}

\author{Davide Rossi}
\orcid{0000-0002-0651-5393}
\email{davide.rossi@unibo.it}
\affiliation{%
  \institution{Universit\`{a} di Bologna}
  \city{Bologna}
  \country{Italy}
}

\author{Francesco Conti}
\orcid{0000-0002-9887-7323}
\email{f.conti@unibo.it}
\affiliation{%
  \institution{Universit\`{a} di Bologna}
  \city{Bologna}
  \country{Italy}
}


\author{Luca Benini}
\orcid{0000-0001-8068-3806}
\affiliation{%
  \institution{ETH Z\"{u}rich}
  \state{Z\"{u}rich}
  \country{Switzerland}
}
\affiliation{%
  \institution{Universit\`{a} di Bologna}
  \city{Bologna}
  \country{Italy}
}

\renewcommand{\shortauthors}{Rogenmoser and Tortorella, et al.}

\thanks{The presented work was supported by Thales Alenia Space.

We acknowledge support by the EU H2020 Fractal project funded by ECSEL-JU grant agreement \#877056.
We acknowledge support by TRISTAN, which has received funding from the Key Digital Technologies Joint Undertaking (KDT JU) under grant agreement nr. 101095947. The KDT JU receives support from the European Union’s Horizon Europe’s research and innovation programmes.

The authors thank Luca Rufer for his valuable contribution to the research project.}

\begin{abstract}
  \input{src/sections/00_Abstract.tex}
\end{abstract}



\received{March 8\textsuperscript{th}, 2023}
\received[revised]{August 15\textsuperscript{th}, 2023}
\received[accepted]{TBD.}

\maketitle

\input{src/sections/01_Introduction.tex}

\input{src/sections/02_Related_Work.tex}

\input{src/sections/03_Architecture.tex}

\input{src/sections/04_Evaluation.tex}

\input{src/sections/05_Discussion.tex}

\input{src/sections/06_Conclusion.tex}

\bibliographystyle{ACM-Reference-Format}
\bibliography{sample-manuscript}











\end{document}

%% file: Commands_Acronyms.tex
\usepackage[nonumberlist,nogroupskip]{glossaries}
\usepackage{siunitx}
\usepackage{multirow}
\usepackage{adjustbox}
\usepackage{pifont}
\usepackage{cleveref}
\usepackage{caption}
\usepackage{subcaption}
\usepackage{threeparttable}
\usepackage{listings}
\usepackage[inline]{enumitem}
\usepackage{tikz,pgfplots,pgfplotstable}
\usepackage{wheelchart} 
\usepackage[normalem]{ulem}
\pgfplotsset{compat=1.17}
\pgfplotsset{
    name nodes near coords/.style={
        every node near coord/.append style={
            name=#1-\coordindex,
            alias=#1-last,
        },
    },
    name nodes near coords/.default=coordnode
}
\tikzset{lines/.style={draw=none},}

\newcommand{\cmark}{\textcolor[HTML]{3E8441}{\ding{51}}}
\newcommand{\xmark}{\textcolor[HTML]{B7352D}{\ding{55}}}
\newcommand{\tmark}{\textcolor[HTML]{E59955}{\textbf{\texttildelow}}}

\newacronym{dmr}{DMR}{Double Modular Redundancy}
\newacronym{tmr}{TMR}{Triple Modular Redundancy}
\newacronym{hmr}{HMR}{Hybrid Modular Redundancy}
\newacronym{soc}{SoC}{System on Chip}
\newacronym{isa}{ISA}{Instruction Set Architecture}
\newacronym{set}{SET}{Single Event Transient}
\newacronym{seu}{SEU}{Single Event Upset}
\newacronym{pulp}{PULP}{Parallel Ultra-Low Power}
\newacronym{dsp}{DSP}{Digital Signal Processing}
\newacronym{tcdm}{TCDM}{Tightly-Coupled Data Memory}
\newacronym{dcls}{DCLS}{Dual-Core Lockstep}
\newacronym{tcls}{TCLS}{Triple-Core Lockstep}
\newacronym{dmode}{DCLS-mode}{Dual-Core Lockstep Mode}
\newacronym{tmode}{TCLS-mode}{Triple-Core Lockstep Mode}
\newacronym{radhard}{Rad-Hard}{Radiation-Hardened}
\newacronym{ecc}{ECC}{Error-Correcting Codes}
\newacronym{pc}{PC}{Program Counter}
\newacronym{rf}{RF}{Register File}
\newacronym{csr}{CSR}{Control and Status Register}
\newacronym{rhbd}{RHBD}{Radiation Hardening by Design}
\newacronym{rhbp}{RHBP}{Radiation Hardening by Process}
\newacronym{dma}{DMA}{Direct Memory Access}
\newacronym{fsm}{FSM}{Finite State Machine}
\newacronym{sp}{SP}{Stack Pointer}
\newacronym{cps}{CPS}{Cyber-Physical System}
\newacronym{scps}{S-CPS}{Space Cyber-Physical System}
\newacronym{cots}{COTS}{Commercial Off-the-Shelf}
\newacronym{ppa}{PPA}{performance, power, and area}
\newacronym{odrg}{ODRG}{On-Demand Redundancy Grouping}
\newacronym{cfft}{CFFT}{Complex Fast Fourier Transform}
\newacronym{pdk}{PDK}{Process Design Kit}

\newcommand{\rebuttal}[1]{#1}

\newcommand{\st}[1]{} 

\definecolor{PULPred}{HTML}{A8322C}
\definecolor{PULPblue}{HTML}{1269B0}
\definecolor{PULPgreen}{HTML}{168638}
\definecolor{PULPorange}{HTML}{F29545}
\definecolor{PULPpurple}{HTML}{910569}

\definecolor{InnovusPink}{HTML}{FF7BF4}
\definecolor{InnovusBlue}{HTML}{00C3FF}
\definecolor{InnovusYellow}{HTML}{FFD500}
\definecolor{InnovusTan}{HTML}{FADDAD}
\definecolor{InnovusRed}{HTML}{990000}
\definecolor{InnovusBrown}{HTML}{E26100}
\definecolor{InnovusGreen}{HTML}{008200}
\definecolor{InnovusHMR}{HTML}{008E8C}

%% file: src/sections/00_Abstract.tex
Space Cyber-Physical Systems (S-CPS) such as spacecraft and satellites strongly rely on the reliability of onboard computers to guarantee the success of their missions.
Relying solely on radiation-hardened technologies is extremely expensive, and developing inflexible architectural and microarchitectural modifications to introduce modular redundancy within a system leads to significant area increase and performance degradation.
To mitigate the overheads of traditional radiation hardening and modular redundancy approaches, we present a novel Hybrid Modular Redundancy (HMR) approach, a redundancy scheme that features a cluster of RISC-V processors with a flexible on-demand dual-core and triple-core lockstep grouping of computing cores with runtime split-lock capabilities.
Further, we propose two recovery approaches, software-based and hardware-based, trading off performance and area overhead.
Running at \SI{430}{\mega\hertz}, our fault-tolerant cluster achieves up to \SI{1160}{\mega OPS} on a matrix multiplication benchmark when configured in non-redundant mode and 617 and 414 MOPS in dual and triple mode, respectively.
A software-based recovery in triple mode requires 363 clock cycles and occupies \SI{0.612}{\milli\meter\squared}, representing a 1.3\% area overhead over a non-redundant 12-core RISC-V cluster.
As a high-performance alternative, a new hardware-based method provides rapid fault recovery in just 24 clock cycles and occupies \SI{0.660}{\milli\meter\squared}, namely $\sim$9.4\% area overhead over the baseline non-redundant RISC-V cluster.
The cluster is also enhanced with split-lock capabilities to enter one of the available redundant modes with minimum performance loss, allowing execution of a mission-critical portion of code when in independent mode, or a performance section when in a reliability mode, with <400 clock cycles overhead for entry and exit.
The proposed system is the first to integrate these functionalities on an open-source RISC-V-based compute device, enabling finely tunable reliability vs. performance trade-offs.

\glsresetall

%% file: src/sections/01_Introduction.tex
\section{Introduction}\label{section:Introduction}
\glspl{cps} integrate real-time computing and communication capabilities with monitoring and control actions over components in the physical world~\cite{shi_survey_2011}. To face the harshness of the space environment, modern space systems such as satellites and spacecraft require tight coupling between onboard processing, communication (cyber), sensing, and actuation (physical) elements~\cite{klesh_cyber-physical_2012}. The orbit determination and control subsystems on a small spacecraft or in satellites' constellations provide a clear link between onboard processing and sensing elements of the spacecraft's physical environment~\cite{di_mascio_-board_2021}, becoming increasingly critical as small spacecraft become ever more capable~\cite{klesh_cyber-physical_2012}.
Radiation-induced soft errors such as \glspl{set} and \glspl{seu} can occur more frequently in space than at ground level, creating the need for additional hardware to mitigate detrimental effects on the system~\cite{wachter_survey_2019}.
In this scenario, digital computing systems representing the decisional part of a \gls{scps} must be designed to be reliable and tolerate faults induced by cosmic radiation.

Various solutions exist to protect electronics from the adverse effects of radiation~\cite{wachter_survey_2019}. \Glspl{scps} typically rely on \glsreset{rhbd}\gls{rhbd} techniques to add reliability at the technology level, yet scaling of these approaches lags behind the scaling of their commercial counterparts, significantly impacting \gls{ppa} of these designs. Insulating techniques~\cite{alles_radiation_2011}, and polymer shielding~\cite{shahzad_views_2022} can also help mitigate soft errors. Furthermore, it is also possible to enhance the fault tolerance capabilities of digital systems by introducing redundancy at different levels in their design flow. Temporal redundancy techniques rely on repeated executions of the same work to determine the correct result~\cite{feng_shoestring_2010}. Spatial or modular redundancy techniques rely on multiple hardware blocks executing the same task and comparing the results~\cite{ginosar_survey_2012}. These approaches rely on rigid schemes for repetition in space and time of redundant blocks or tasks. Hence, they can severely impact the \gls{ppa} of computing platforms. This leads to a significant gap between \glspl{soc} for space and their commercial counterparts, which make use of technological advances unencumbered by radiation-induced faults.

Therefore, the increasing demand for strong processing capabilities in space~\cite{xie_resource-cost-aware_2018} is pushing researchers toward lower-overhead solutions and use of more advanced technologies to close this gap in \gls{ppa}. In recent years, the advent of RISC-V and open-source hardware has encouraged the development of high-performance \glspl{soc} for various domains without licensing or other restrictions. This includes the space domain, where custom modifications to improve properties such as reliability~\cite{di_mascio_open-source_2021} and fault tolerance are often required. Among proposed architectures, heterogeneous systems with multi-core computing clusters have gained traction in the space industry~\cite{ginosar_rc64_2016} due to increased performance and flexibility for computation and \gls{dsp} workloads~\cite{kurth_hero_2018}.
While multiple processors offer increased performance for parallelizable tasks, they also provide a unique opportunity for reliability enhancements: multiple cores can execute identical tasks, comparing their results to detect and react to faults.

In this paper, we introduce a multi-core RISC-V-based computing system for space featuring an innovative \gls{hmr} approach. We leverage the independent cores available in a multi-core RISC-V cluster built on a commercial technology for redundant execution in a dynamically runtime configurable manner, supporting on-demand reconfiguration in \gls{dcls} and \gls{tcls} modes. This extends the \gls{odrg} presented in~\cite{rogenmoser_-demand_2022}, which presents a \gls{tcls} configurable cluster with software-based recovery and configuration prior to startup.
Our design extends this by allowing each application to configure its reliability setting according to its requirements, possibly decided at runtime.
Furthermore, we implement two recovery alternatives, software and hardware-assisted, comparing their impact on the hardware resources and performance in case of a fault. The checking, voting, and switching hardware in the implemented design does not affect the internals of the processor core, allowing for the use of verified RISC-V processor cores without requiring any internal (potentially erroneous) modifications to rapidly build a reliable system. What we propose is, to the best of our knowledge, the most compact, flexible, and configurable computing cluster offering the best trade-off between performance and reliability.

Besides extending our previous work~\cite{rogenmoser_-demand_2022}, we introduce the following novel contributions:
\begin{itemize}
    \item A re-configurable computing cluster for Dual-Core Lockstep and Triple-Core Lockstep execution capable of tackling compute-intensive and safety-critical applications at the same time. The proposed cluster can be configured so that the computing cores can operate independently  during high-performance execution or in Dual/Triple-Core Lockstep mode in case of safety-critical tasks.
    \item Robust hardware extension for fast fault recovery. We introduced dedicated Error-Correcting Codes-protected status registers to restore the state of the computing cores to the closest reliable state in time. This feature allows the cores to perform cycle-by-cycle backups of their internal state in the protected registers, reducing by $15\times$ the required time to recover from a fault  over the software-based approach.
    \item A novel runtime-programmable split-lock mechanism, allowing for fast switching and re-configuration between the available redundant modes. With these features, it is possible to explicitly define portions of code within a \textit{mission-critical section}, configuring the cores for safe lockstep execution, or within a \textit{performance section}, disabling the lockstep execution temporarily, with minimum configuration switching overhead.
\end{itemize}

To validate our proposed approach, we implemented the RISC-V cluster in Global Foundries 22~\si{\nano\meter} technology, achieving up to \SI{430}{\mega\hertz} operating frequency and \SI{1160}{\mega OPS} when configured in independent mode and 617 and 414 MOPS in \gls{dmr} and \gls{tmr} mode, respectively. With only software-based recovery features, the proposed cluster occupies \SI{0.612}{\milli\meter\squared} with just 1.3\% area overhead over the non-redundant configuration, featuring 363 clock cycles time-to-recovery in triple mode. When enhanced with hardware-based recovery features, it provides rapid fault recovery in just 24 clock cycles occupying \SI{0.660}{\milli\meter\squared}, $\sim$9.4\% area overhead over the baseline RISC-V cluster. The proposed split-lock mechanism allows for entering and exiting a redundancy mode in <400 clock cycles for mission-critical code execution, or <200 exiting and re-entering the redundancy mode for temporarily more performance.
To foster future research in computer architecture for space, we release the proposed architecture as the first fully open-source~\footnote{https://github.com/pulp-platform/redundancy\_cells} RISC-V-based multi-core cluster with a finely tunable trade-off between reliability and performance.

%% file: src/sections/02_Related_Work.tex
\section{Related Work}\label{section:Related_Work}

\subsection{Design Trends for CPS in Space}

In recent years, \gls{scps} and onboard processors in space have been striving for more performance, fueled by more advanced mission requirements and higher expectations for onboard electronics resulting from the advancement of commercial technology~\cite{furano_ai_2020, reinhardt_high_2019}.
However, reliability and dependability~\cite{xie_resource-cost-aware_2018, bhuiyan_dependability_2018} remain key concerns in the space domain because, with high levels of cosmic rays, errors are more frequent than at ground level and occur too often for a system to perform its intended mission reliably.

To achieve higher levels of performance, the industry at large has steered towards multi-core processing architectures, leveraging multiple parallel processing cores to increase performance. Players in space computing have adopted this approach: systems such as the BAE RAD5500 architecture~\cite{berger_quad-core_2015}, the Gaisler GR740 architecture~\cite{hijorth_gr740_2015}, and the DAHLIA NG-ultra architecture~\cite{danard_ng-ultra_2022} leverage quad-core architectures built on existing, radiation-hardened processes. Other designs, like Ramon Chip's RC64~\cite{ginosar_rc64_2016}, leverage large processor arrays for DSP calculations to improve performance. However, these systems, built on PowerPC, SPARC, ARM, and a custom \gls{isa}, respectively, suffer from high costs due to specialized and limited production and lack of commercial technology and software support, further increasing their cost~\cite{di_mascio_open-source_2021}. On the other hand, the reliability and dependability of these designs built on \gls{radhard} technology increase the chances of mission success, making them the go-to choice for complex, high-cost space missions.

However, not all missions in the space domain require these high tolerance levels. New Space missions with smaller, more cost-effective satellites tend to rely more on \gls{cots} electronics to design the spacecraft~\cite{toorian_cubesat_2008}. These components offer significantly more performance at a lower cost, sacrificing reliability, which can be tolerated for non-critical applications, such as in CubeSats~\cite{toorian_cubesat_2008} or for non-critical machine learning workloads~\cite{furano_ai_2020}. However, even for these satellite missions, certain aspects, such as communication and control, still require some radiation tolerance in \gls{cots} solutions, often guaranteed with watchdog timers.

On the other hand, large-scale infrastructure may have satellites as one part of the pipeline but can contain many different components, from IoT sensor nodes to high-performance server infrastructure. In such cases, using many different components with different programming paradigms can lead to untenable cost overheads, where component reuse can greatly benefit the designers. In such a case, reliability may be required by the satellite but is counter-productive for the power-constrained sensor node requiring high performance and energy efficiency. Furthermore, reusing a single component both for commercial and reliability environments can greatly reduce component cost, leveraging scale to combat the cost issue plaguing the space industry.

One of the recent trends in custom \gls{soc} design is the shift to RISC-V, leveraging a modern and open-source \gls{isa} to develop custom hardware. RISC-V is also quickly impacting the space domain~\cite{di_mascio_open-source_2021, furano_european_2022}, which relies on modified custom designs to ensure proper reliability and fault tolerance. In using this new \gls{isa}, the space industry can benefit not only from designing custom, reliable hardware for their purposes but also from the commercial environment. Companies developing dedicated hardware for space have also adopted this approach: Gaisler is developing the NOEL-V architecture~\cite{andersson_development_2020}, Microchip is leveraging SiFive's processor architectures for NASA's next High Performance Space Computing project~\cite{leibson_nasa_2023}, and dedicated projects like De-RISC~\cite{wessman_-risc_2022} are pushing this development.
In this work, we continue on this trend by exploiting an open-source industrially verified RISC-V core~\cite{gautschi_near-threshold_2017, group_openhw_2023} as a building block for our system.

\subsection{Fault Tolerance Approaches}

\subsubsection{Radiation-Induced Faults in Space}
Since the beginning of spaceflight, the radiation environment has been an object of investigation~\cite{van_allen_observation_1958}, showing to be significantly more severe than on earth, with a flux over \SI{e7}{particles\per\centi\meter\squared\per day}~\cite{bourdarie_near-earth_2008}. The effect on electronics, also essential for spaceflight's development, became apparent shortly after the first space missions~\cite{ziegler_effect_1979}, where high energy ions from cosmic rays cause charge separation within semiconductors, leading to voltage spikes within transistors.
These voltage spikes can lead to inverted bits within memory cells or transient spikes at the output of logic gates. These are called ``soft errors,'' as they can be corrected simply by re-writing the logic value and do not cause permanent faults within the circuit. A further subdivision of soft errors is between \glsreset{set}\glspl{set} for transient events in combinational logic and \glsreset{seu}\glspl{seu} for events affecting state-keeping logic and memories. Both \glspl{set} sampled by a register and \glspl{seu} (directly) can cause incorrect system behavior.
While early integrated circuits seemed tolerant to these adverse effects, the hazardous space environment severely affects the more deeply integrated nodes used onboard newer satellites~\cite{wilkinson_tdrs-1_1991}, leading to a requirement for reliable electronics. For example, on a \SI{65}{\nano\meter} technology node, systems experience on the order of \SI{e-7}{errors\per bit\per day}; this error rate is increasing for smaller technology nodes and is more and more due to \glspl{set} as clock frequencies are increasing~\cite{di_mascio_open-source_2021}. The error rate is also estimated to vary with the flying orbit of the system~\cite{engel_predicting_2006}, leading to a higher probability of a \gls{set}-induced erroneous value being sampled.

\rebuttal{Fault mitigation approaches for ASICs' protection vary from using Radiation-Hardened technologies to architectural and system-level modifications intended to be adopted independently from the chosen technology to software-based approaches for protecting \gls{cots} devices. The state-of-the-art approaches are qualitatively discussed in the following and summarized in Table~\ref{tab:related_work}.}

\subsubsection{Radiation Hardened by Design}
\glsreset{radhard}\gls{radhard} technologies rely on silicon-level techniques, like transistor resizing and low-level design modifications, to increase the robustness of the technology cells towards particles striking the silicon. Industry groups like Gaisler~\cite{andersson_leon_2017, hijorth_gr740_2015}, BAE~\cite{berger_quad-core_2015}, DAHLIA~\cite{danard_ng-ultra_2022}, and Ramon Chips~\cite{ginosar_rc64_2016} mostly rely on \gls{radhard} technologies to increase the fault-resilience of space-grade multi-core processors. This increases the designs' fault tolerance capabilities by reducing the engineering effort in implementing architectural solutions for application-specific \glspl{soc}. However, one of the negative aspects of \gls{radhard} cells is that they are not readily available, being significantly more expensive than standard technological processes. Moreover, \gls{radhard} technologies, while significantly more resilient towards soft errors than commercial technologies, are not completely immune~\cite{di_mascio_open-source_2021}. To integrate the tolerance features, \gls{radhard} cells are significantly larger than those used in standard technologies due to their resilience requirements. Radiation hardening requires major investments when moving to a new technology node, both in time and money. Therefore, \gls{radhard} libraries and design kits are generally only available for older technology nodes, significantly lagging behind commercial counterparts. This leads to increased costs due to custom, low-volume designs for low-performance and low-efficiency designs.  While this is acceptable for small, mission-critical systems, it is problematic if applications require higher computing performance on a spacecraft.


\subsubsection{Architectural Modifications}
Rather than relying on technology-level hardening, a processor's architecture can also be used to enhance a system's reliability. This can mitigate one of the key performance and efficiency concerns, allowing modern technology nodes to be used for the design, albeit requiring more advanced architectural considerations and design. The two common approaches are \begin{enumerate*}[label=(\roman*)]
\item \gls{ecc} for information redundancy of the stored bits (for example, in registers or memories) and
\item modular redundancy, namely \glsreset{dmr}\gls{dmr} and \glsreset{tmr}\gls{tmr}, at various levels in the design, ensuring that multiple redundant copies of a critical module process the same information and produce the same outcome.
\end{enumerate*}

\Gls{ecc} is one of the most common and efficient ways to protect static data in \glspl{soc}' memories and registers. The encoding and decoding logic allows for individual data words' protection with limited additional bits, resulting in far less overhead than modular redundancy approaches. It is often used to further bolster designs protected by \gls{radhard} technologies, improving their reliability. Other examples include the Duck-Core~\cite{li_duckcore_2021}, which implements \gls{ecc} in the pipeline registers, rolling back to allow re-execution of the last instruction in case a fault occurs.

When considering modular redundancy, the size of the replicated module considerably impacts the design effort, functionality, and performance. STRV~\cite{walsemann_strv_2023} implements \gls{tmr} at a very fine granularity within a RISC-V core, replicating the circuitry and voting after each register to ensure correct processing. Similarly, \citeauthor{gkiokas_fault-tolerant_2019}~\cite{gkiokas_fault-tolerant_2019} propose a \gls{tmr} approach inside a 5-stage dual issue time predictable core triplicating the fetch, decode, and execute stage of the pipeline. The results provided by the three execution stages are propagated to a voter in charge of deciding which should continue in the pipeline to the memory and register write-back stages.

The two approaches, \gls{ecc} and modular redundancy, can also be combined well. SHAKTI-F~\cite{gupta_shakti-f_2015} uses a hybrid approach, using \gls{ecc} for registers and memory while implementing \gls{dmr} for the ALU within the processor's execution stage, ensuring the calculation is executed correctly. While these modifications can significantly increase reliability, they require severe architectural modification of the underlying component and unrecoverable area overhead. Furthermore, Duck-Core and SHAKTI-F focus only on data protection within the cores, leaving the problem of control-flow faults uncovered.

The final architectural modifications to consider are full replications of the component blocks, using modular redundancy at a far coarser granularity. Prime examples are \glsreset{dcls}\gls{dcls} and \glsreset{tcls}\gls{tcls}, replicating the entire processing core and adding checkers and voters at the boundary. These system-level solutions are closer to the approach proposed in this work and are detailed further below in \Cref{sec:system-rel}.

\subsubsection{Software Approaches}
Reliability can also be tackled in software, leveraging existing hardware \gls{cots} and adding fault tolerance in a later step. 
One option is to leverage multiple threads, duplicating~\cite{alcaide_software-only_2020} or triplicating~\cite{barbirotta_design_2022} an application on separate threads executing concurrently or sequentially. These multiple executions are matched with a final checking step, either in hardware or software, to ensure correct execution. While these approaches require little to no modification of the underlying hardware, they incur a heavy penalty in performance, either through additional time for execution or the use of multiple available threads, limiting parallelism. Performance is further penalized by the checking step, requiring either time or dedicated hardware resources to check for correctness. While this approach can properly handle faults in the data path, some faults in the processor's internal control logic may lead to an unrecoverable fault, possibly stalling a thread, thereafter requiring a watchdog timer to restart all processing properly with an extreme penalty.

Another software mitigation strategy is re-execution, which heavily uses \gls{cots} hardware and temporal reliability. Shoestring~\cite{feng_shoestring_2010} duplicates instructions to protect only those segments of code that can result in user-visible faults when subjected to a software error without first showing faulty behavior. Similarly, Inherent Time Redundancy~\cite{reddy_inherent_2007} consists in re-executing a program instruction with varying inputs. The assumption behind this approach is that certain microarchitectural events in a processor execution depend only on the instructions and not on the provided inputs. \citeauthor{jagtap_software_2020}~\cite{jagtap_software_2020} use system reset with a watchdog timer to re-execute software that experienced a fault. Contrarily to hardware redundancy, it is possible to implement software-based approaches without introducing area overhead, as the replication of hardware components that have to be checked or voted on is unnecessary. Checking or voting can be performed in software but remains vulnerable to errors. Furthermore, re-execution and software checking introduces non-negligible processing overhead, impacting the performance of the protected application.

\subsection{System-level Reliability Methods}\label{sec:system-rel}
System-level redundancy approaches make it possible to increase the reliability of digital systems without affecting the internal architecture of vulnerable components. ARM Cortex-R processors~\cite{arm_cortex-r5_2011} and other ARM offerings feature a \glsreset{dcls}\gls{dcls} option~\cite{iturbe_soft_2016}, allowing the \gls{soc} to group two cores within the design. Furthermore, this implementation allows for re-configuration of the architecture while the cores are in reset, switching from a locked operation to a split one. The architecture does not specify the behavior in case an error occurs, relying on the surrounding \gls{soc} to determine fallback behavior and to handle and repair any remaining faults, generally with a reset or relying on check-pointing.
However, critical applications on \gls{cps} with tight timing constraints may require faster repair mechanisms or even continuous operation, not tolerating any downtime due to an error that occurred. Onboard computers in spacecraft are responsible for executing critical real-time operations, which might tolerate minimum execution time variation to avoid the mission's failure~\cite{zhang_fault_2003}. Thus, ensuring that the system can satisfy strict time requirements during the recovery from incurring faults is essential.

Error recovery procedures for a \gls{dcls} implementation can be quite complex, as it is not possible to determine the correct output in case of mismatch. Therefore, \citeauthor{iturbe_arm_2019}~\cite{iturbe_arm_2019} implemented a \glsreset{tcls}\gls{tcls} design with the Cortex-R5 processor. The approach relies on an \textit{assist unit} that wraps the cores to group them in triplets so that all the grouped cores can operate in lockstep with identical inputs and outputs decided with majority voting. The \textit{assist unit} relies on \gls{radhard} technology, while the remaining components use commercial technologies and are reinforced with modular redundancy and \gls{ecc}. Any fault in a single core can be repaired during operation as the two remaining cores continue processing. A re-synchronization routine of the core ensures that any errors remaining in its state bits can be corrected. The cores store all the required internal state information in system memory, are reset by the \textit{assist unit}, and read their stored state again from memory. While not discussed in detail in~\cite{iturbe_arm_2019}, we assume the re-configuration behavior of ARM \gls{tcls} to be similar to the \gls{dcls} implementation, requiring a reset to switch between a split and a locked state.
Another common processing core for reliable systems is the AURIX TriCore~\cite{infineon_aurixtricore_2016}, built on a custom \gls{isa} for embedded applications. This processor also relies on a \gls{tcls} implementation to detect and correct faults, requiring a reset to recover from latent errors.

CEVERO~\cite{silva_cevero_2020} is an example of \gls{dcls} implementation using RISC-V-based Ibex~\cite{davide_schiavone_slow_2017} cores. Unlike ARM, it relies on a hardware-based recovery mode that consists in restoring their internal state, namely the \gls{pc} and \gls{rf}, to a known safe one. Thus, CEVERO implements a \textit{safety island} containing a backup copy of the cores' \gls{pc} and \gls{rf}, replicated every time the cores have matching outputs. In case of a mismatch, the two cores roll back to the last saved status. This is done by resetting the two cores, copying the backup copy of the \gls{pc} and \gls{rf} back into the cores, and restarting their operation. This solution introduces additional area overhead due to the backup registers. On the other hand, it requires only 40 clock cycles for a rollback to complete, significantly increasing the performance of the rollback procedure compared to other SW-based approaches. In this design, the cores in lockstep cannot be decoupled, creating a strong trade-off between area overhead and performance. In addition, the \gls{pc} and \gls{rf} alone do not represent the complete state to which a core should recover, leading to incomplete recovery. Moreover, no protection mechanism is applied to ensure the reliability of the backup registers.

\citeauthor{shukla_low-overhead_2022}~\cite{shukla_low-overhead_2022} implement a re-configurable \gls{dcls} approach within a quad-core processor based on the RISC-V \gls{isa}. The simple, custom-designed core is paired with a mode control unit and a fault detection and correction unit, which ensure proper execution and correction depending on a dedicated mode selection signal. If a fault is detected, the program counter is held to repair the fault. While this may correct any error during execution, an error in the \gls{rf} might remain uncorrected. 

\citeauthor{kempf_adaptive_2021}~\cite{kempf_adaptive_2021} propose a dual-core adaptive runtime-selectable lockstep based on a LEON processor. While normally operating in independent mode, a master core will request the slave core's assistance with a critical task requiring reliability. The slave core halts its execution and assists the master core with the critical task, requiring a pipeline flush to start proper lockstep execution. The register file is not affected, as this is assumed to be \gls{ecc} protected, and only the master core's register file is used. The LEON core is extended with a commit stage between the execute and memory stages to compare the results of the lockstepped cores' execution. If an error occurs, the checker does not commit the instruction and flushes the pipeline to re-execute the failed instruction. This implementation offers an adaptive lockstep mechanism with fast hardware-based entry and exit switching, but it does not fully protect the core, focusing only on the data flow. Furthermore, deep modifications to the LEON core's architecture are still needed, as direct access to the registers of the main core and the ALU of the slave core is required, and an additional pipeline stage is added.

SafeDE~\cite{bas_safede_2021}, SafeDM~\cite{bas_safedm_2022}, and \citeauthor{marcinek_variable_2023}~\cite{marcinek_variable_2023} propose system-level approaches that enforce the concept of diversity in the execution of redundant threads. In particular, SafeDE enforces diversity by running redundant threads on different cores but introduces limitations in terms of applicability. It cannot be used to ensure the safe execution of parallelizable applications or for applications that require IO accesses. On the other hand, SafeDM allows a wider applicability range. However, the two proposed solutions only mitigate the effect of common cause failures. \citeauthor{marcinek_variable_2023} propose a variable-delayed dual-core lockstep approach to avoid common mode failure, but with no possibility to decouple the two cores for independent execution.

The work presented in this paper belongs to the hardware redundant approaches. It proposes the implementation of system-level hardware enhancements such as independent/\gls{dcls}/\gls{tcls} split-lock, and \gls{ecc}-protected backup status registers for fast re-configuration, re-synchronization, and recovery of a multi-core computing cluster. Furthermore, as detailed in \Cref{section:Architecture}, our approach relies on using unmodified industrially verified cores, therefore guaranteeing fault protection without requiring re-verification of the protected cores from a functional viewpoint.

\input{src/tables/relwork_table}

%% file: src/tables/relwork_table.tex
\begin{table}[t]
    \centering
    \caption{Qualitative comparison of related work with corresponding features and configuration.}
    \begin{adjustbox}{width=1\textwidth,center=\textwidth}
    \begin{tabular}{@{}cl crccc@{}}\toprule
         & System & \gls{isa} & Cores & Reliability Method & Configurable & Development \\ \midrule
        \multirow{4}{*}{\rotatebox[origin=c]{90}{\gls{radhard}}} & Gaisler GR740~\cite{hijorth_gr740_2015} & SPARC & 4 & \gls{radhard} \& \gls{ecc} & \xmark & Commercial \\
         & BAE RAD5500~\cite{berger_quad-core_2015} & PowerPC & 4 & \gls{radhard} & \xmark & Commercial \\
         & DAHLIA~\cite{danard_ng-ultra_2022} & ARM & 4 & \gls{radhard} & \xmark & Commercial \\
         & Ramon Chips RC64~\cite{ginosar_rc64_2016} & CEVA X DSP & 64 & \gls{radhard} & \xmark & Commercial \\
        \midrule
        \multirow{4}{*}{\rotatebox[origin=c]{90}{Arch.}} & Duck-Core~\cite{li_duckcore_2021} & RISC-V & 1 & \gls{ecc} & \xmark & Research \\
         & STRV~\cite{walsemann_strv_2023} & RISC-V & 1 & gate-level \gls{tmr} & \xmark & Research \\
         & \citeauthor{gkiokas_fault-tolerant_2019}~\cite{gkiokas_fault-tolerant_2019} & RISC & 1(3) & block-level \gls{tmr} & \xmark & Research \\
         & SHAKTI-F~\cite{gupta_shakti-f_2015} & RISC-V & 1 & block-level \gls{dmr} \& \gls{ecc} & \xmark & Research \\
        \midrule
        \multirow{10}{*}{\rotatebox[origin=c]{90}{System-level}} & ARM \gls{dcls}~\cite{arm_cortex-r5_2011, iturbe_soft_2016} & ARM & 2 & \gls{dcls} & \tmark & Commercial \\
         & ARM \gls{tcls}~\cite{iturbe_arm_2019} & ARM & 3 & \gls{tcls} & \tmark & Research \\
         & AURIX TriCore~\cite{infineon_aurixtricore_2016} &  TriCore & 3 & \gls{tcls} & \tmark & Commercial \\
         & CEVERO~\cite{silva_cevero_2020} & RISC-V & 2 & \gls{dcls} & \xmark & Research \\
         & \citeauthor{shukla_low-overhead_2022}~\cite{shukla_low-overhead_2022} & RISC-V & 4 & \gls{dcls} & \cmark & Research \\
         & \citeauthor{kempf_adaptive_2021}~\cite{kempf_adaptive_2021} & SPARC & 2 & \gls{dcls} \& \gls{ecc} & \cmark & Research \\
         & SafeDE~\cite{bas_safede_2021} & RISC-V & 2 & diverse redundancy & \cmark & Research \\
         & SafeDM~\cite{bas_safedm_2022} & RISC-V & 2 & diverse redundancy & \cmark & Research \\
         & \citeauthor{marcinek_variable_2023}~\cite{marcinek_variable_2023} & RISC-V & 2 & \gls{dcls} & \xmark & Research \\
         & \textbf{This work} & \textbf{RISC-V} & \textbf{12} & \textbf{\gls{dcls} \& \gls{tcls} \& \gls{ecc}} & \textbf{\cmark} & \textbf{Research} \\
         \bottomrule
    \end{tabular}
    \end{adjustbox}
    \label{tab:related_work}
\end{table}

%% file: src/sections/03_Architecture.tex
\section{Architecture}\label{section:Architecture}

The hardware template we rely on is the multi-core \gls{pulp} cluster~\cite{rossi_pulp_2015}, which can be extended with an external host subsystem featuring a controller core, a larger memory, and interfaces for external peripherals. The host subsystem is not area and performance-critical, and we assume it can be made fault-tolerant by means of high-overhead techniques.

\subsection{Background: The PULP Cluster}\label{sec:pulp_cluster}
The \gls{pulp} cluster~\cite{rossi_pulp_2015} features a parametric number of 32-bit CV32E40P cores~\cite{gautschi_near-threshold_2017}, enhanced for fast \gls{dsp} calculations and industrially verified by the OpenHW Group~\cite{group_openhw_2023}.
Each core features an instruction interface connected to a hierarchical instruction cache~\cite{chen_scalable_2023} consisting of one private bank per core, each configured to store \SI{512}{B}. Each private bank fetches instructions from a larger, shared cache of \SI{4}{KiB}, improving the performance of applications using the Single Program Multiple Data (SPMD) paradigm.

In addition, each core features a data interface that connects it to the rest of the system through dedicated demultiplexers for direct access to the system's \gls{tcdm} and all other memory-mapped peripherals. The \gls{tcdm} comprises a parametric number of \SI{32}{bit} word-interleaved memory banks featuring single-cycle access latency. A banking factor of 2 (i.e., the number of memory banks is twice the number of cores) is typically used to minimize the memory banks contention probability, guaranteeing data sharing overhead smaller than 5\%, even for highly memory-intensive workloads. If a collision occurs, round-robin arbitration guarantees fairness and avoids starvation.

Along with the connection to the \gls{tcdm}, the cores have access to both memory-mapped devices within the cluster and the host domain through a peripheral interconnect. One of the core-local peripherals is the \gls{dma} unit, capable of up to \SI{64}{bit\per cycle} data transfers in either direction, full duplex between the external larger memories in the host subsystem and the cluster's local \gls{tcdm}.

Finally, the cores directly connect to an event unit~\cite{glaser_energy-efficient_2021} responsible for synchronization barriers within the cluster. Each core attempts to read from the corresponding register within the event unit and only receives a response once all cores request the same barrier address. Furthermore, the event unit manages interrupts within the cluster, masking and forwarding incoming interrupt signals to the responsible core.

To ensure the cluster can easily access the host system, the cluster has both an input and an output AXI port connected to an AXI interconnect. Through this interconnect, a host can access the cluster's internal memory and peripherals for configuration. To ensure proper operation, the instruction cache, the \gls{dma}, and the cores through the peripherals interconnect have access to the host system's memory.

The \gls{pulp} ecosystem offers open-source software for parallel code execution on the multi-core cluster~\cite{montagna_streamlining_2021}. For various host systems, the software is provided to configure and boot the accelerating cluster and basic low-level drivers that allow for parallel software execution. The software ecosystem also provides a variety of parallel applications and benchmarks, enabling fast development of custom workloads. 

\subsection{On-Demand Redundancy Grouping}
The baseline of the work described in this paper is the \glsreset{odrg}\gls{odrg}~\cite{rogenmoser_-demand_2022}, which consists in a wrapper that allows for \gls{tcls} grouping of the cluster cores. We have extended the \gls{odrg} by also allowing \gls{dcls} grouping and implementing hardware extensions for fast fault recovery and re-synchronization of the redundant cores. Furthermore, we extended the previous work with a faster split-lock mechanism that allows for rapid runtime-selectable reconfiguration of the cluster. This allows the execution of dedicated portions of code that can be defined as critical in one of the available redundant modes while operating in individual mode, or vice versa, a non-critical portion of code while operating in a reliable mode.

\begin{figure}[t]
    \centering
    \includegraphics[width=0.95\columnwidth]{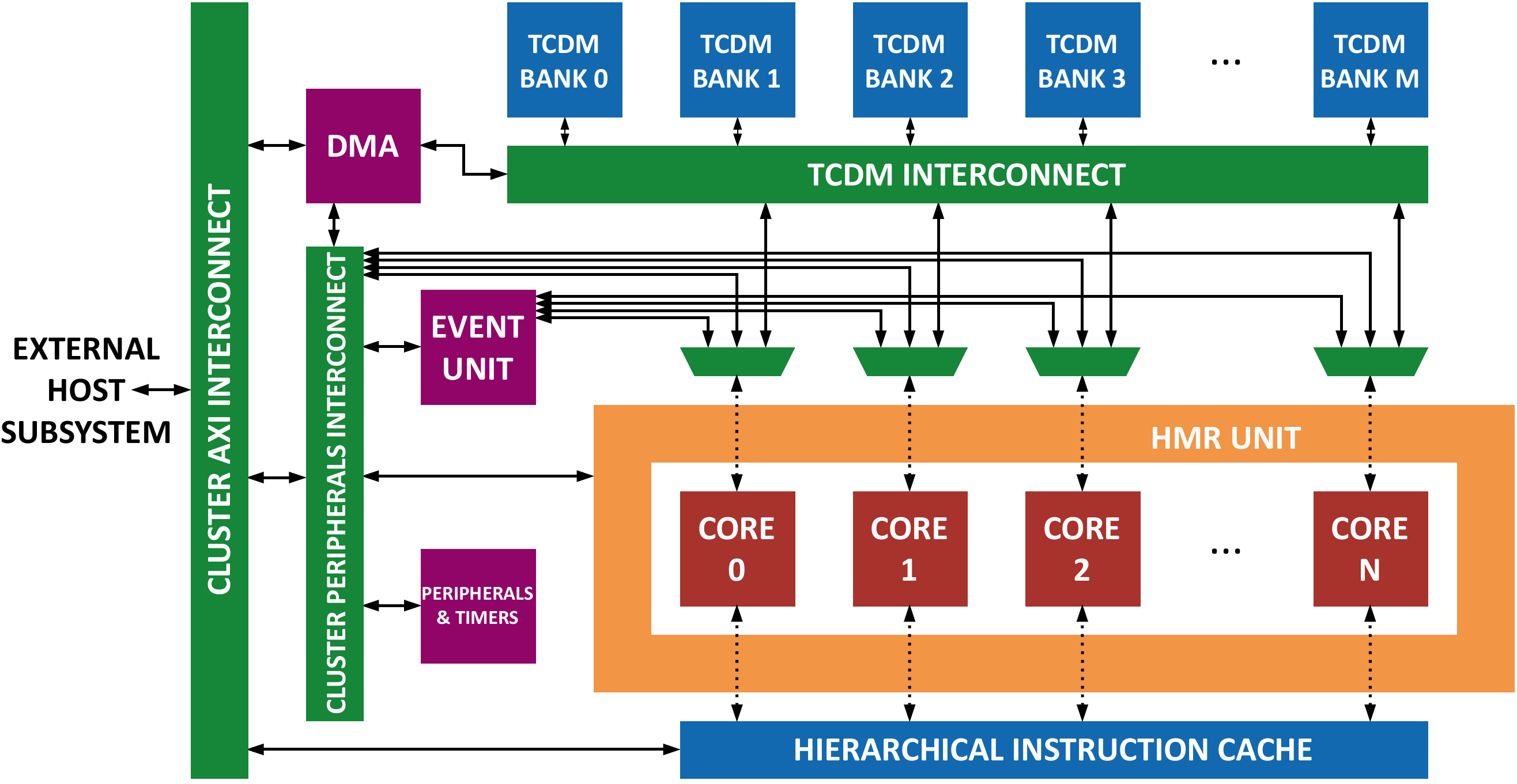}
    \caption{Integration of the \gls{hmr} unit within the \gls{pulp} cluster.}
    \label{fig:hmr}
\end{figure}

\newpage

\subsection{System-Level Integration}\label{sec:integration}
We focus on implementing modular redundancy within the \gls{pulp} cluster to safeguard the individual processor cores' correct operation. For this purpose, our design approach and experimental evaluations assume that the modules outside the cores, particularly the \gls{tcdm} and the instruction memories, are reliable and protected by dedicated safety-critical techniques, such as \gls{ecc} encodings and memory scrubbers.
In the \gls{hmr} cluster, the cores are grouped into lockstep, either in \gls{dcls} or \gls{tcls}. All inputs and outputs to the cores pass through a \gls{hmr} unit surrounding the cluster cores, as shown in \Cref{fig:hmr}, where the \gls{hmr} unit is integrated in the \gls{pulp} cluster described in \Cref{sec:pulp_cluster}. In a \gls{pulp} cluster with 12 cores, the \gls{hmr} unit can easily divide all cores into groups of 2 and 3 for \gls{dcls} and \gls{tcls}, respectively, without remainder. Should a different number of cores be chosen, the respective mode will be unavailable for the remaining cores.

To allow access to its configuration registers, the \gls{hmr} unit exposes a peripherals memory port connected to the peripherals interconnect in the \gls{pulp} cluster. Through this interconnect, any core within the cluster and any controller outside the cluster with access to the host memory port can configure the \gls{hmr} unit. Furthermore, this memory port also allows the readout of all other memory-mapped registers within the unit, such as the current reliability state and error statistics measured within the unit.

\subsection{Hybrid Modular Redundancy Block}\label{sec:hmr_block}

Modular redundancy requires multiple hardware instances performing the same operation to ensure correct execution, with the result being compared in a \gls{dmr} case or voted in a \gls{tmr} case. For reliable calculations, all inputs of two or three cores are directly connected with each other, ensuring that the cores receive identical signals for subsequent processing. The outputs of the cores are then connected to dedicated checkers or majority voters to ensure matching calculations or detect any faults that might have occurred and then apply a recovery. When grouped, the locked cores behave within the system as a single virtual core.

While the cores are protected in case a fault occurs, the checkers and majority voters may also be vulnerable to soft errors. To mitigate this, we assume the cores implement additional protections for the bus, for example using \gls{ecc}, being checked on a protocol level within the protected core region. Thus, any error occurring within the checkers and majority voters will result in a bus error, which can then be corrected.


\begin{figure}[t]
    \centering
    \includegraphics[width=0.95\columnwidth]{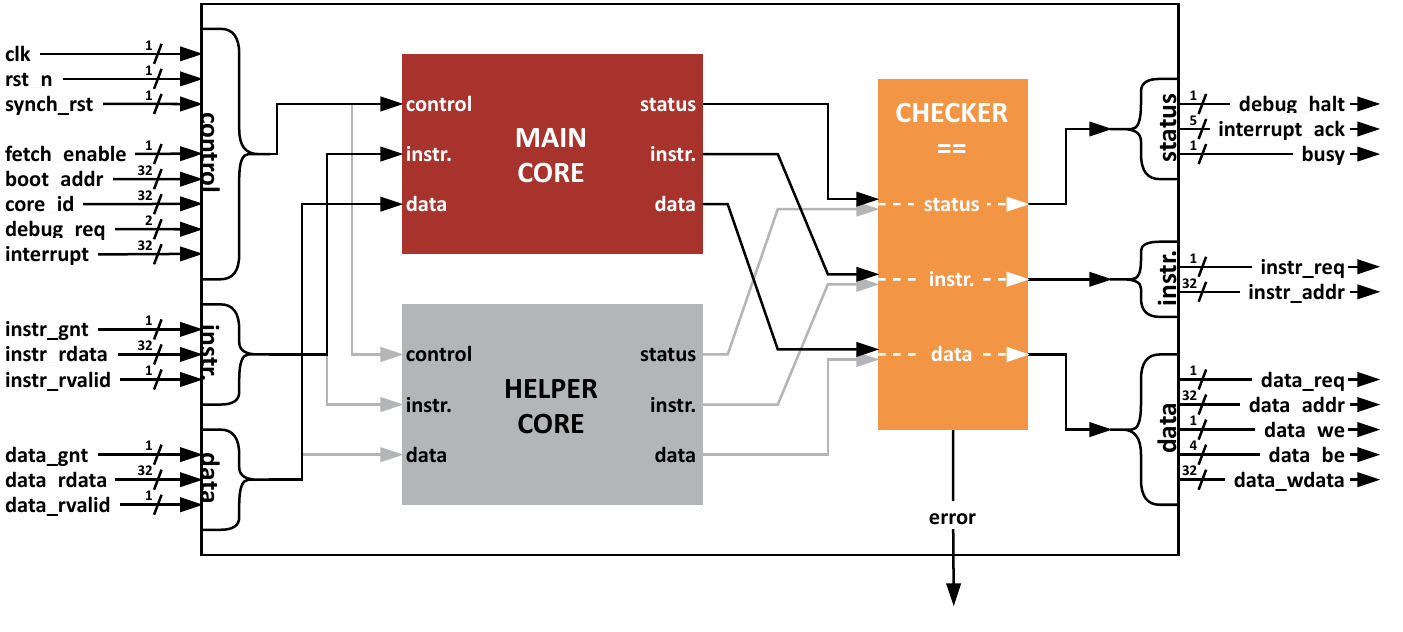}
    \caption{\acrfull{dmode} grouped cores.}
    \label{fig:dmr}
\end{figure}

\subsubsection{Dual-Core Lockstep}\label{subsection:DCLS}
When configured in \gls{dmode}, the cores in the \gls{pulp} cluster are paired, identifying one main core and one helper core, as shown in \Cref{fig:dmr}. From a system perspective, one core is disabled to no longer act on the system's interconnect, while the other continues processing based on both cores' execution.

The cores of each group are configured to receive the same input information, so they are expected to generate the same output results. The outputs of each core in a group are then propagated to a bitwise checker that detects whether the results produced by the two cores are different. If the results of the two cores match, the checker selects only the result of the main core as an effective output of the pair, propagating it to the rest of the system. On the other hand, in case of a mismatch, the checker raises an error signal, indicating that a fault affected the status of one of the two cores in the group. Since it is impossible to know which core produced the correct output in \gls{dmode}, the checkers gate their outputs toward the system to avoid fault propagation. Furthermore, the error is directly signaled to the system to start a recovery procedure.

\begin{figure}[t]
    \centering
    \includegraphics[width=0.95\columnwidth]{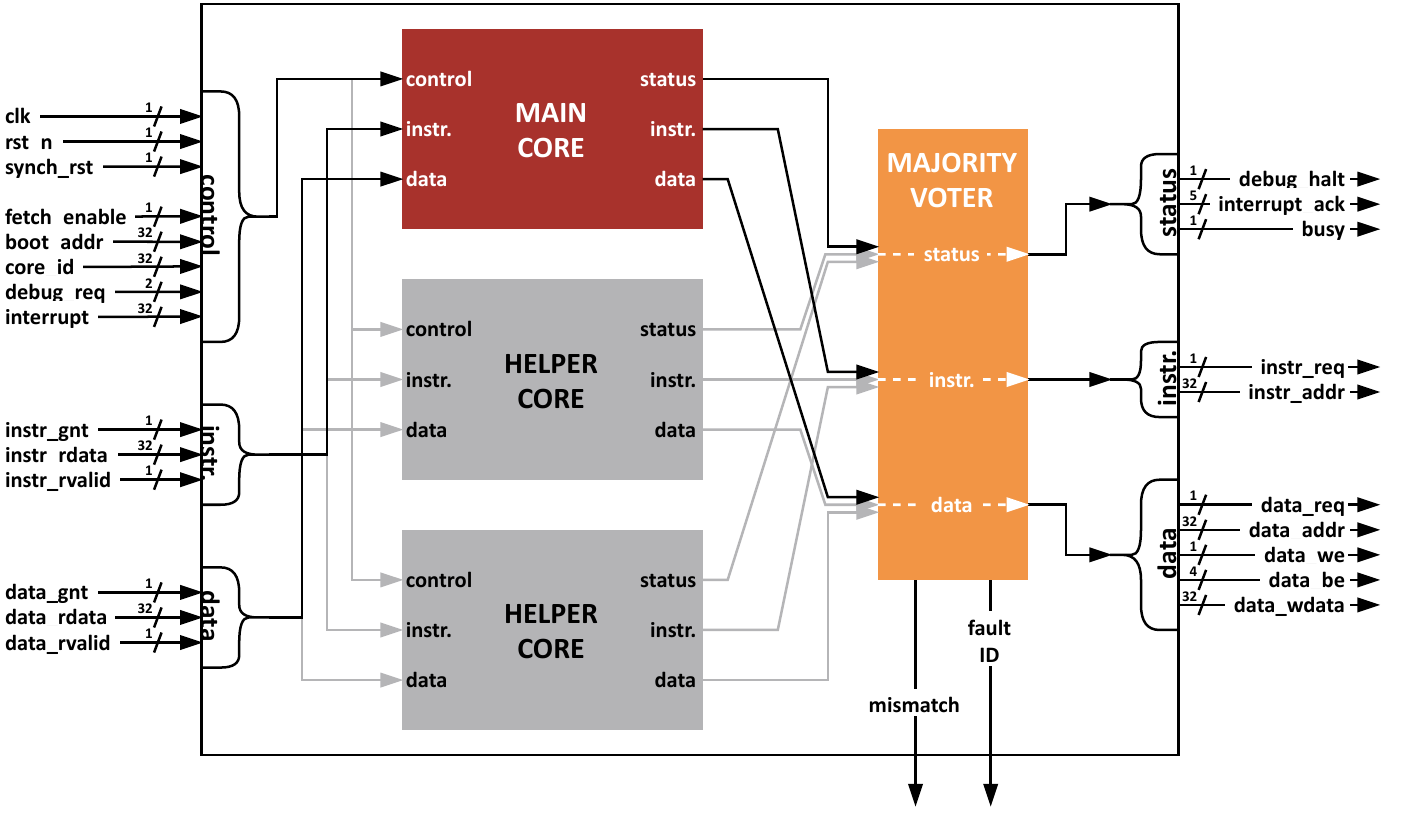}
    \caption{\acrfull{tmode} grouped cores.}
    \label{fig:tmr}
\end{figure}

\subsubsection{Triple-Core Lockstep}\label{subsection:TCLS}
In the \gls{tmode}, three cores are grouped, identifying one main core and two helper cores, as shown in \Cref{fig:tmr}. As in \gls{dmode}, the inputs from the system are shared among the cores in the group, ensuring that they operate on identical data and control signals. However, unlike in \gls{dmode}, the outputs of the cores are connected back to the system through bitwise majority voters. Logic to disable the connection is unnecessary, as the voter properly selects the correct output. Each majority voter compares the outputs from the grouped cores and raises an error signal only if it detects a mismatch between them. Unlike \gls{dmode}, since it is possible to vote over the results of three cores in \gls{tmode}, the state of the faulty core can be restored using the state of the other two non-faulty cores. While not advisable with high error rates, letting the two non-faulty cores continue their operation in lockstep without correction can be enabled within the \gls{hmr} unit, delaying the faulty core's re-synchronization. However, if a second mismatch is detected, the grouped cores must enter a re-synchronization routine immediately.

\subsubsection{Parametrization}
While the \gls{hmr} unit offers both \gls{dmode} and \gls{tmode} configurations, the hardware can be parametrized to disable these different modes or enforce them permanently. This means that the \gls{hmr} unit can offer either only \gls{dmode}, only \gls{tmode}, or both, supporting configuration switching with the split-lock explained below. If a desired instantiation does not require split-lock functionality, \gls{dmode} or \gls{tmode} can be enforced permanently. However, these configurations were not investigated in detail.

The \gls{hmr} unit is designed to be agnostic to the cores, requiring proper configuration to adjust to the core's interface. We tested it with the CV32E40P~\cite{gautschi_near-threshold_2017} core, the default for the \gls{pulp} cluster and the configuration used here, and the Ibex~\cite{davide_schiavone_slow_2017} core.

\subsection{Split-Lock}
\begin{figure}[t]
    \centering
    \begin{subfigure}[c]{0.9\columnwidth}
         \centering
         \includegraphics[width=\textwidth]{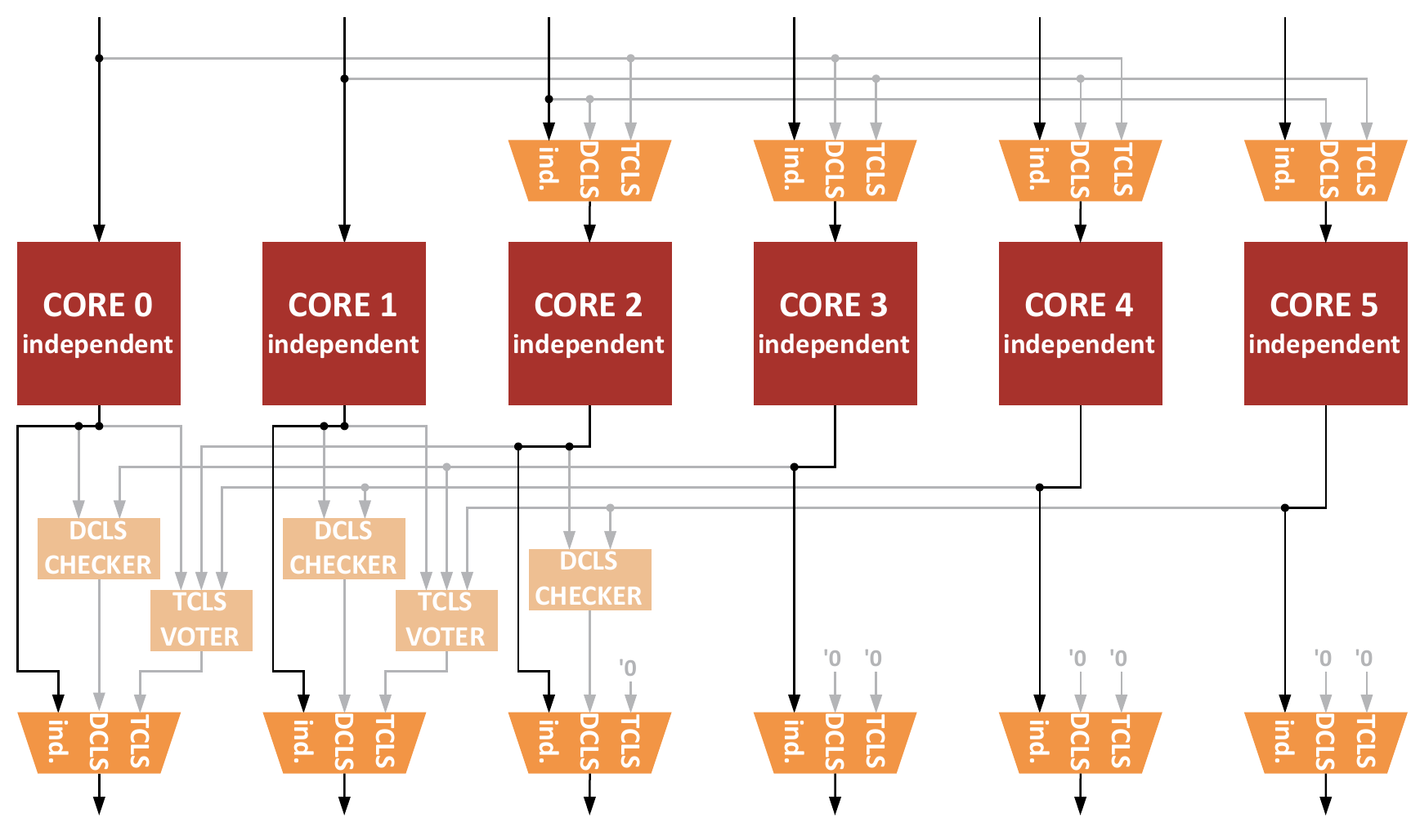}
         \caption{Split-Lock in independent mode}
         \label{fig:split-lock-independent}
     \end{subfigure}
     \begin{subfigure}[c]{0.9\columnwidth}
         \centering
         \begin{subfigure}[t]{0.45\textwidth}
             \includegraphics[width=\textwidth]{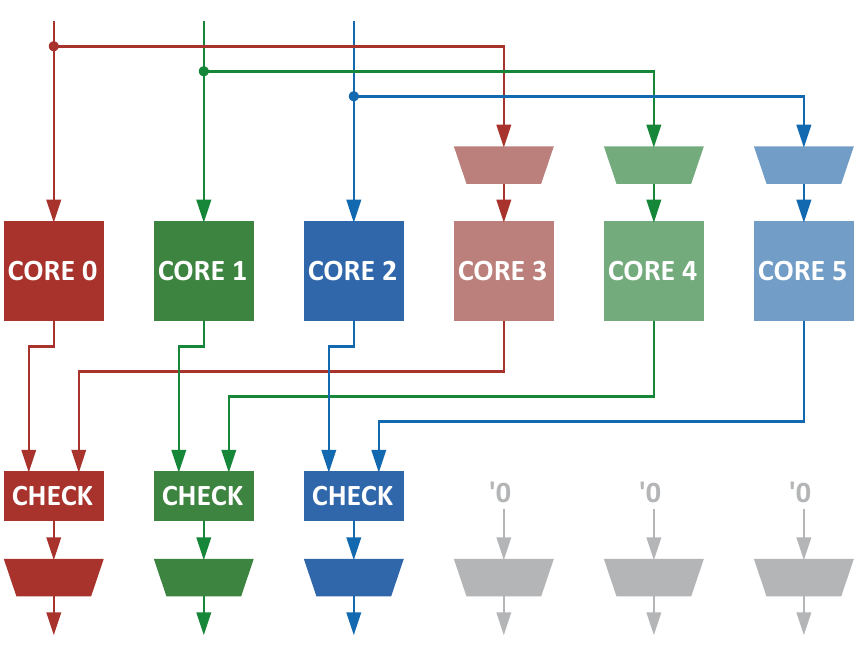}
             \caption{Split-Lock in \gls{dmode}}
             \label{fig:split-lock-dcls}
         \end{subfigure}
         \hfill
         \begin{subfigure}[t]{0.45\textwidth}
             \includegraphics[width=\textwidth]{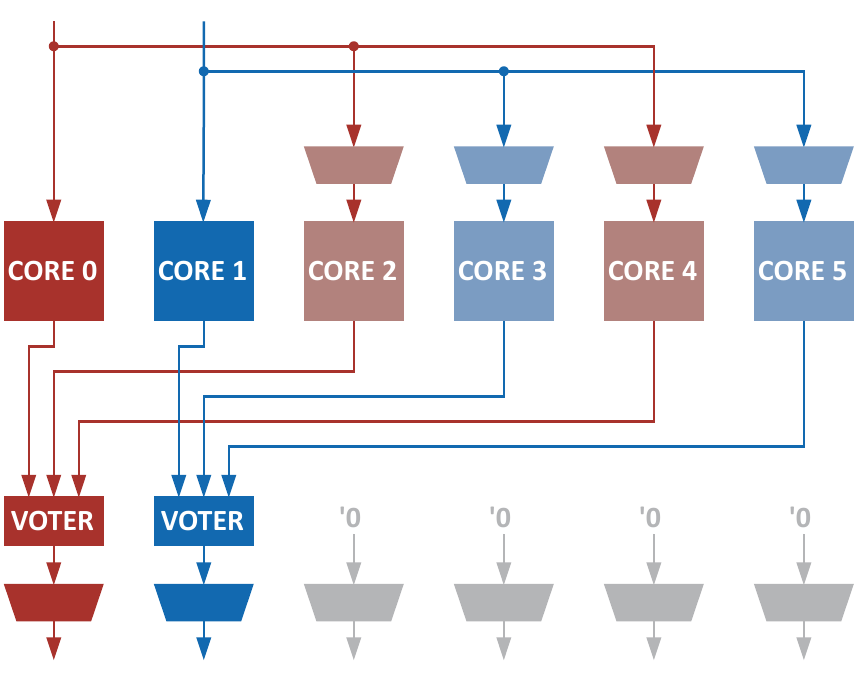}
             \caption{Split-Lock in \gls{tmode}}
             \label{fig:split-lock-tcls}
         \end{subfigure}
     \end{subfigure}
    \caption{Split-Lock muxing example for a 6-core group configuration.}
    \label{fig:split-lock}
\end{figure}

Locking cores together into redundant configurations permanently is limiting for some application domains that might tolerate the occurrence of faults and require higher performance. For example, a satellite must satisfy tighter resilience constraints during orbital maneuvers than during image processing for the satellite's payload. Thus, the proposed \gls{pulp} cluster can be configured before startup to work in the independent mode, \gls{dmode}, or \gls{tmode}, following the grouping procedure shown in \Cref{fig:split-lock}. This allows any application running on the host subsystem to configure the cluster for its purposes and reliability requirements after a cluster reset and before its boot sequence, locking the execution for the accelerated application into the specified reliability mode.

The main differentiator of the cores in the cluster is their respective identifier (ID). It starts at zero and is incremented for each additional core, allowing for easy identification of work sections in a parallelized task. To keep this convenience when the reliability modes are enabled, the \gls{dmode} and \gls{tmode} grouping of the cores is performed in an interleaved fashion, as shown in the 6-cores example depicted in \Cref{fig:split-lock}. \Cref{fig:split-lock-dcls} (\gls{dmode}) shows that core 0 is grouped with core 3, while in \Cref{fig:split-lock-tcls} (\gls{tmode}), it is grouped with core 2 and core 4. This results in preserving the lowest core IDs when enabling all cores in the \gls{dmode} or \gls{tmode} reliability configurations, keeping a simple parallelization software scheme. Furthermore, each reliability group can be configured independently, allowing some cores to execute a reliable application without impeding the others from operating in independent mode.

Changing the reliability mode is also possible at runtime, allowing the cores to switch configurations based on the demands of the software application executed thereon. When starting in the individual configuration, the application can explicitly declare a \textit{mission-critical section} with a portion of code that must be executed reliably. Alternatively, when starting in the reliable configuration, the application can declare a \textit{performance section} with a portion of code that does not require stringent reliability guarantees. This allows a system to make use of the tradeoffs between reliability and performance on a single \gls{soc} with minimal configuration overheads.


\subsubsection{Mission-Critical Section}\label{sec:missioncriticalsection}
The \textit{mission-critical section} is a temporary reliable state of execution where two or three cores operate in lockstep, while otherwise remaining in the independent configuration. It is designed such that all cores assisting the main core for reliable execution only pause their own thread, returning to it after the \textit{mission-critical section} completes. As the cores' internal registers (\glsreset{pc}\glsreset{rf}\glsreset{csr}\gls{pc}, \gls{rf}, and \glspl{csr}) are used during the reliable execution, the helper cores' state is temporarily saved to the stack in the \gls{tcdm}, allowing them to retrieve it later.

\lstset{
  frame=single,
  language=C,
  basicstyle=\scriptsize,
}
\begin{figure}[b]
\vspace{-0.2cm}
\noindent\begin{minipage}{.50\textwidth}
\begin{lstlisting}[language=C, caption=Mission-Critical Section Wrapping Function, label=lst:critical]
int mission_critical_sec(int (*fn_handle)()) {
    enable_lockstep();
    // Interrupt locks cores together
    int ret = fn_handle();
    disable_lockstep();
    return ret;
}
\end{lstlisting}
\end{minipage}
\hfill
\begin{minipage}{.42\textwidth}
\begin{lstlisting}[language=C, caption=Mission-Critical Section Entry Interrupt Service Routine, label=lst:entry]
void synchronization_irq(void) {
    store_state_to_stack();
    store_pc_to_reg(core_id());
    barrier();
    // Cores are locked together
    // Synchronous Clear Possible
    reload_pc_and_state(core_id());
    asm volatile ("mret");
}
\end{lstlisting}
\end{minipage}
\end{figure}

To enter the \textit{mission-critical section}, the main core executes a custom software routine, shown in Listing~\ref{lst:critical}, that writes the desired state (\gls{dmode} or \gls{tmode}) into the configuration register within the \gls{hmr} unit. 
This triggers an interrupt in the cores that have to participate in the group, indicating they must start the execution of a \textit{mission-critical section}. 
To ensure the cores are awake and available to serve it, the interrupt is routed through the cluster's event unit waking up the cores even if they are asleep waiting for a barrier.
The interrupt service routine outlined in Listing~\ref{lst:entry} stores the internal state of the cores onto the stack in the \gls{tcdm}. The internal state is represented by 30 modifiable registers of the \glsreset{rf}\gls{rf}, excluding the \gls{sp}, and by specific \glspl{csr}, such as the \textit{MEPC}, which stores the last \glsreset{pc}\gls{pc} executed before that interruption.

Once the entire state is stored onto the stack, the \gls{sp} is stored in a memory-mapped register within the \gls{hmr} unit, indicating that this \textit{unload} phase is complete, and the \textit{reload} phase can start.
After storing the state, the cores enter a synchronization barrier and are locked together, ensuring they all continue as one.
Here, it is also possible to trigger a synchronous clear towards the locked cores to bring the internal flip-flops of the cores to their default value without polluting the reset network.
As the \gls{sp} storage register is linked to the core's ID, the grouped cores that share the same ID after grouping read the \gls{sp} register and load the state back from the stack in parallel. After completion, the interrupt service routine is exited into the \textit{mission-critical section} outlined by the software.

To exit the \textit{mission-critical section}, the locked cores write the new desired state (i.e., independent mode) to the \gls{hmr} unit configuration registers. This write operation frees all the cores from their locked state, so the main core can continue its execution
, as its internal state is still consistent. On the other hand, the helper cores must return to the state before the \textit{mission-critical section} entry. Therefore, the \gls{hmr} module synchronously clears the helper cores bringing them back to the boot sequence, where they use their memory-mapped \gls{sp} register to reload their state back from the \gls{tcdm}.

\subsubsection{Performance Section}

The \textit{performance section} allows an application typically requiring reliable execution to forego the stringent requirements and temporarily leverage the higher parallel performance possible with independent cores. It is designed to allow the main thread to continue processing, with the assisting cores branching from the main thread temporarily, after which the temporary threads are resolved at the end of the performance section.

To enter the \textit{performance section}, the cores split apart by a single write to a configuration register in the \gls{hmr} unit. While the main core can continue the main thread, the newly separated cores simply update their \gls{sp} to an independent region and also continue the application. Differentiated based on the core's ID, the separated cores operate on their own stack to avoid interfering with the main thread's stack. 

Once the \textit{performance section} is complete, the main core stores its internal state, and all cores enter a synchronization barrier to be re-grouped. The state of the separated cores is abandoned, as the section is only intended to be temporary to provide additional performance. Once locked together after the synchronization barrier, the cores load the main core's state in lockstep, similar to the entry of the \textit{mission-critical section} described above, and continue the application.

\subsection{Fault Recovery}
In the \gls{dmode} and \gls{tmode} outlined in \Cref{sec:hmr_block}, a fault would not immediately impact the system as the outputs of the cores are either gated in \gls{dmode} or voted in \gls{tmode}. However, further execution may not be possible, as the outputs remain gated in the \gls{dmode}, and any additional errors in another core of the group in \gls{tmode} may cause complete failure. To avoid this scenario, the possibly corrupted state within the cores must be corrected, and the locked cores must be re-synchronized. It is also essential to guarantee that the system recovery can be performed in time to avoid critical tasks from failing. On-board computers execute critical tasks that must be guaranteed to complete in very limited time intervals~\cite{zhang_fault_2003}. Consequently, if incoming faults corrupt the system's status, ensuring a minimal time to recovery to meet real-time operations' constraints is fundamental.


\subsubsection{TCLS-Mode Software Re-synchronization Routine}\label{sec:sw-resynch}
When an error occurs in \gls{tmode} affecting a single core, the cluster can continue operating, as the voters overrule the invalid outputs of the erroneous core. Furthermore, in \gls{tmode}, the three cores provide redundant internal state information, allowing the system to recover to a fully functional and correct state without additional hardware. This implies storing the internal state of one of the non-faulty cores into the stack placed in TCDM during the \textit{unload} stage and then reloading the state back to all three cores during the \textit{reload} stage, similar to the \textit{mission-critical section}.

\begin{figure}[t]
    \centering
    \includegraphics[width=0.7\columnwidth]{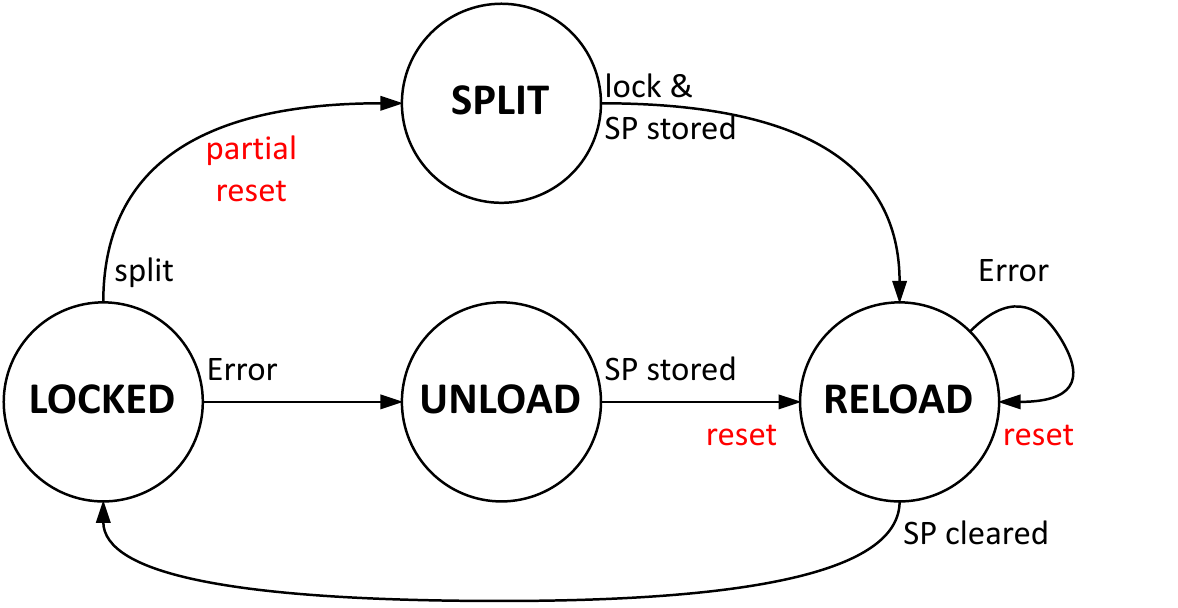}
    \caption{\gls{fsm} of the \gls{tmode} controller.}
    \label{fig:tcls_fsm}
\end{figure}

To do this, as shown in \Cref{fig:tcls_fsm}, the \gls{tmode} \gls{fsm} of the \gls{hmr} unit enters the \textit{unload} state and sends an interrupt to the cores through the event unit, triggering an interrupt service routine. This routine stores the internal state of the non-faulty cores (i.e., \gls{pc}, \gls{rf}, and \glspl{csr}) into the stack. During the execution of the \textit{unload} routine by all three cores, the bitwise voter scans the contents of these cores and stores only the state of the non-faulty cores in memory. Then, the \gls{sp} is stored in the same memory-mapped register mentioned in the \textit{mission-critical section} which is located within the \gls{hmr} unit. This indicates to the \gls{tmode} \gls{fsm} that the \textit{unload} stage is complete, allowing it to enter the \textit{reload} state. On this transition, the \gls{tmode} \gls{fsm} can also trigger a synchronous clear toward the locked cores if configured.

To refill the cores with the correct state, the cores use the memory-mapped \gls{sp} register to reload the main core's state stored in memory back into the locked cores. Furthermore, the cores can be cleared again if another fault happens during the \textit{reload} phase, avoiding the execution of another \textit{unload} operation since a safe state was already stored in memory. To properly execute this routine even after a clear, the boot sequence of the cores directly checks the \gls{sp} register to see if the stored value is not zero. If its content is not zero, that value is used by the cores as a reload \gls{sp}. On the other hand, if the register content is zero, the cores continue with a normal boot sequence. Therefore, once the reload is complete, the software stores back a zero value to the \gls{sp} register in the \gls{hmr} unit, switching the \gls{tmode} \gls{fsm} back into a nominal \textit{run} state. The cores then execute an \texttt{mret} instruction, returning to the application running before the fault recovery interrupt was triggered.

\subsubsection{Hardware-Based Re-synchronization and Rapid Recovery Extension}
\begin{figure}[t]
    \centering
    \includegraphics[width=\columnwidth]{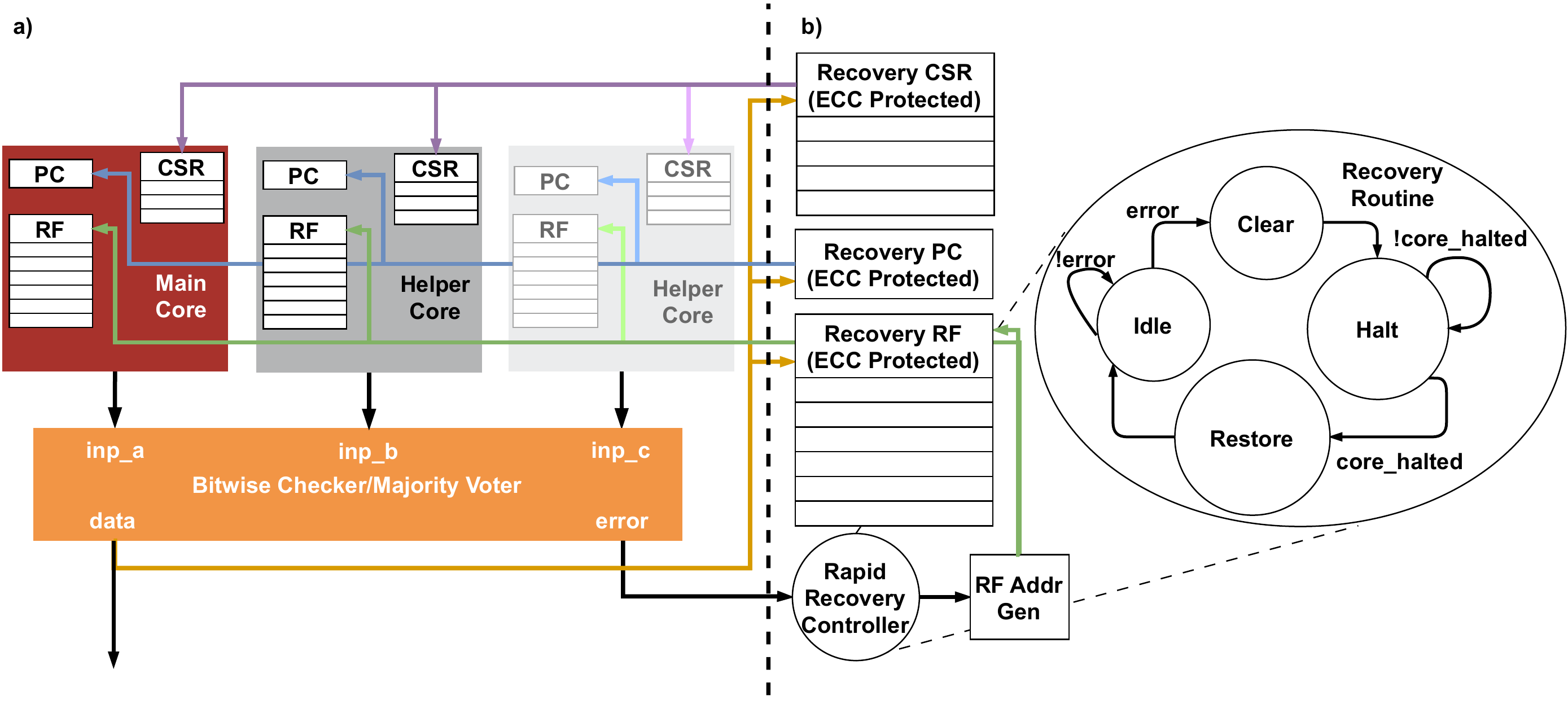}
    \caption{Implementation of the \textit{rapid recovery} hardware extension, highlighting a) the grouped cores, and b) the \textit{rapid recovery} region containing the ECC-protected status registers and the finite state machine of the \textit{rapid recovery} routine.}
    \label{fig:rapidrecovery}
\end{figure}

The software-based \textit{unload} and \textit{reload} procedures execution can be further improved by integrating dedicated hardware for \textit{rapid recovery} with minor modifications to the processor core's architecture. As in the software-based recovery mechanism, \textit{rapid recovery} relies on the fact that the state of each core is defined by its \gls{pc}, \gls{rf}, and \glspl{csr}. Instead of copying the state of the cores to memory and exploiting a software-based interrupt service routine, we extended the \gls{hmr} unit by introducing a hardwired recovery engine for each group of cores. \Cref{fig:rapidrecovery}a shows the cores grouped in \gls{dmode} or \gls{tmode}, connected to a bitwise checker or a majority voter, as described in \Cref{sec:hmr_block}. The \textit{rapid recovery} region, depicted in \Cref{fig:rapidrecovery}b, comprises four main modules: recovery \gls{pc}, recovery \gls{rf}, recovery \glspl{csr}, and a \textit{rapid recovery} controller.

During normal processing without any errors, a backup of the main core's \gls{pc}, \gls{rf}, and \glspl{csr} content is copied into the registers of the \textit{rapid recovery} region. This operation is done each cycle, as the core's interface was modified to expose the write ports of \gls{pc}, \gls{rf}, and \glspl{csr} to propagate them to the backup registers. Hence, the backup write operation is accomplished in a single cycle. Furthermore, the recovery \gls{pc}, recovery \gls{rf}, and recovery \glspl{csr} are protected with internal \gls{ecc} encoding/decoding, configurable through parameter selection at design time.

If the checkers or voters detect a mismatch, they raise an error signal that blocks write operations to the registers in the \textit{rapid recovery} region, ensuring their content is not corrupted by a faulty state. The same error signal triggers the \textit{rapid recovery} controller that starts the recovery routine, depicted in \Cref{fig:rapidrecovery}b. The error flag from the checker or voter sends the \gls{fsm} of the \textit{rapid recovery} routine from the \textit{Idle} state to the \textit{Clear} state. This transition raises a synchronous clear signal to the faulty group, which brings the internal flip-flops of the cores to their default value without polluting the reset network. Then, the recovery routine jumps into the \textit{Halt} state, where a debug request signal is raised towards the faulty cores, forcing them to enter the debug mode. Shortly after they receive the debug request, the cores raise the halt response signal toward the \textit{rapid recovery} controller. In debug mode, the cores are halted and allow the recovery hardware to access their internal registers without interference.

Once the faulty cores have raised the halt response signal, the recovery routine jumps into the \textit{Restore} state. In this state, the \gls{pc}, \gls{rf}, and \glspl{csr} content of the faulty cores is reloaded from the recovery \gls{pc}, recovery \gls{rf}, and  recovery \glspl{csr}. This process is executed sequentially, restoring the content of all 31 modifiable registers of the \gls{rf}, exploiting its two write ports in parallel. The \gls{pc} and the \glspl{csr} are reloaded in parallel to the \gls{rf}. 

The same hardware introduced to enable the \textit{rapid recovery} can be used to speed up the process of entering a \textit{mission-critical section} and exiting the \textit{performance section}. The recovery \gls{pc}, recovery \gls{rf}, and recovery \glspl{csr} are used to back up the state of the main core continually so that this state can be used to enter the routine safely. As the helper cores software is designed to be executed in order in the \textit{mission-critical section}, they still require their state to be saved to the stack in the TCDM. Therefore, the main core immediately enters the barrier once the interrupt occurs, while the helper cores store their state to the stack and enter the barrier afterward. After all the cores have entered the barrier and are locked together, the \textit{rapid recovery} hardware executes its recovery routine, synchronizing the state of the locked cores to continue into the \textit{mission-critical section} or return to the reliable mode after the \textit{performance section}.

%% file: src/sections/04_Evaluation.tex
\section{Evaluation}\label{section:Evaluation}

\begin{figure}[t]
    \centering
    \begin{subfigure}[c]{0.55\textwidth}
        \centering
        \includegraphics[width=0.8\textwidth]{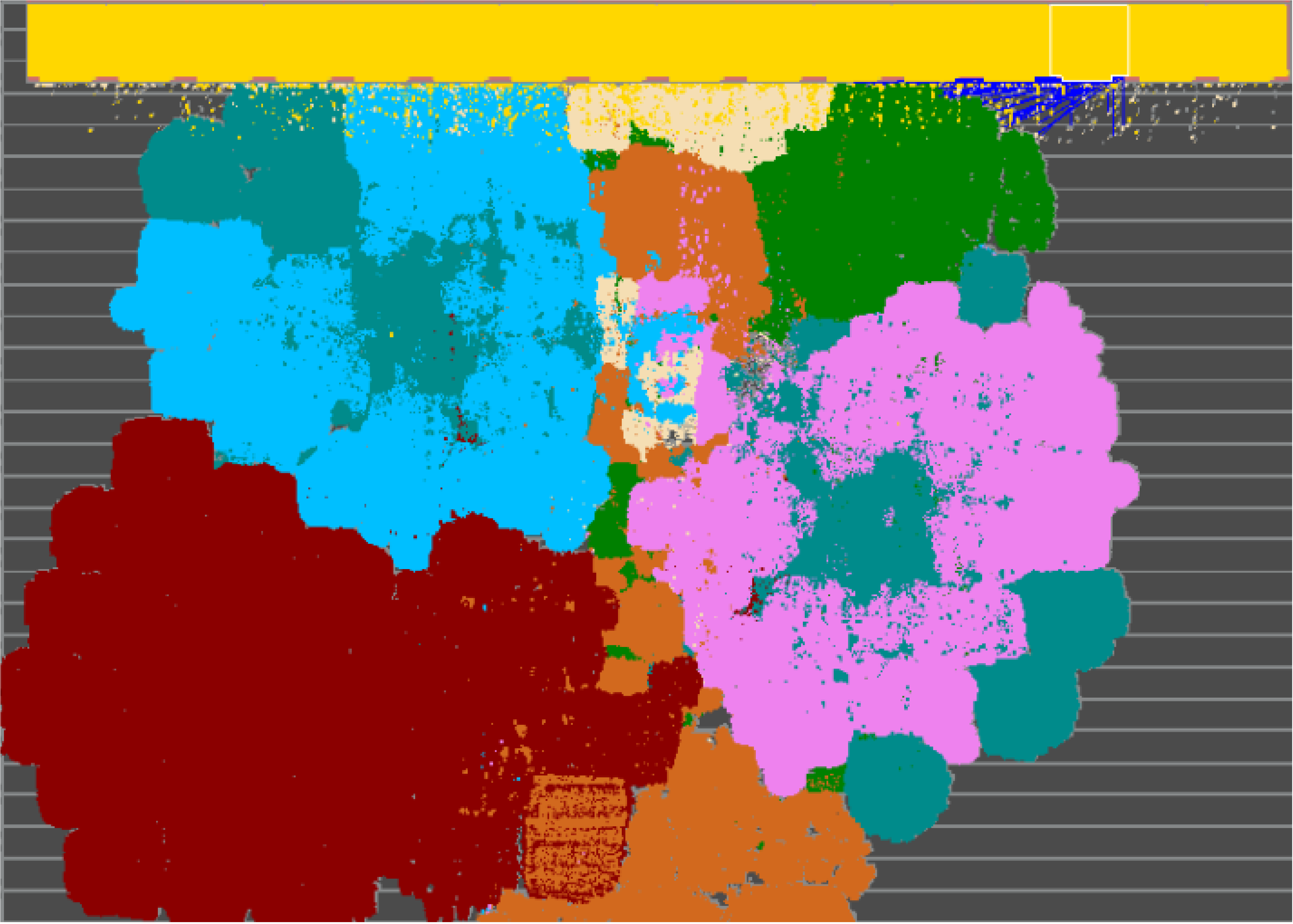}
        \caption{Layout of the physical implementation of the fault-tolerant \gls{pulp} cluster featuring full \gls{hmr} and \textit{rapid recovery}.}
        \label{fig:layout}
    \end{subfigure}
    \hfill
    \begin{subfigure}[c]{0.4\textwidth}
        \centering
        \caption{\gls{pulp} cluster area comparison in all available configurations.}
        \begin{tabular}{@{}lrr@{}}\toprule
            \multicolumn{2}{@{}l}{PULP Cluster Area [\si{\milli\meter\squared}]} & Overhead \\ \midrule
            Baseline & 0.604 & - \\
            DMR & 0.605 & 0.3\% \\
            TMR & 0.608 & 0.7\% \\
            HMR & 0.612 & 1.3\% \\ \midrule
            \multicolumn{3}{@{}c@{}}{With Rapid Recovery} \\ \midrule
            DMR & 0.654 & 8.4\% \\
            TMR & 0.657 & 8.8\% \\
            HMR & 0.660 & 9.4\% \\
            \bottomrule
        \end{tabular}
        \label{tab:area}

    \end{subfigure}
    \vspace{0.2cm}
    \begin{subfigure}[c]{0.60\columnwidth}
        \centering
        \begin{tikzpicture}
            \def\WCtest#1#2{
                \ifdim \WCpercentage pt>5 pt
                    #1
                \else
                    #2
                \fi
            }
            \wheelchart[
                radius={1}{2},
                start angle=-10,
                explode=\WCvarD,
                data={\WCtest{\WCvarC}{\WCvarC~(\WCperc)}},
                data angle shift=\WCvarE,
                counterclockwise,
                lines = 0.5,
                lines ext =0.1,
                lines sep =-0.3
            ]{
                9/InnovusHMR/{\textbf{\gls{hmr} Unit}}/0.6/0/white/,
                15/InnovusPink/{Cores (Odd IDs)}/0/0/black/,
                15/InnovusBlue/{Cores (Even IDs)}/0/-10/black/,
                18/InnovusYellow/{\gls{tcdm}}/0/-20/black/,
                28/InnovusRed/{Instruction Cache}/0/25/white/,
                2/InnovusTan/{\gls{tcdm} Interconnect}/0/0/black/,
                5/InnovusGreen/{\gls{dma}}/0/0/black/,
                8/InnovusBrown/{AXI Interconnect}/0/0/black/
            }
            \wheelchart[
                radius={1}{2},
                start angle=-10,
                explode=\WCvarD,
                data={},
                wheel data={\WCtest{\WCperc}{}},
                wheel data style={text=\WCvarF},
                wheel data pos=0.4,
                counterclockwise,
                slices style={fill=none}
            ]{
                9/InnovusHMR/{\textbf{\gls{hmr} Unit}}/0.6/0/white/,
                15/InnovusPink/{Cores (Odd IDs)}/0/0/black/,
                15/InnovusBlue/{Cores (Even IDs)}/0/-10/black/,
                18/InnovusYellow/{\gls{tcdm}}/0/-25/black/,
                28/InnovusRed/{Instruction Cache}/0/25/white/,
                2/InnovusTan/{\gls{tcdm} Interconnect}/0/0/black/,
                5/InnovusGreen/{\gls{dma}}/0/0/black/,
                8/InnovusBrown/{AXI Interconnect}/0/0/black/
            }
        \end{tikzpicture}
        \caption{Area breakdown of the fault-tolerant \gls{pulp} cluster featuring full \gls{hmr} and \textit{rapid recovery}, with the same color as in the physical implementation layout.}
        \label{fig:pie_full}
    \end{subfigure}
    \begin{subfigure}[c]{0.4\columnwidth}
        \centering
        \begin{tikzpicture}
            \wheelchart[
                data={\WCvarC~(\WCperc)},
                data angle shift=\WCvarD,
                radius={0.6}{1.2},
                start angle=-10,
                lines =0.5,
                lines ext=0.1,
                lines sep = -0.2
            ]{
                78/PULPblue!80/{Rapid Recovery}/-80,
                7/PULPorange!80/{Control}/0,
                5/PULPgreen!80/{DMR Check}/0,
                10/PULPred!80/{TMR Vote}/0
            }
        \end{tikzpicture}
        \caption{\gls{hmr} unit area breakdown.}
        \label{fig:pie_hmr}
    \end{subfigure}
\hfill
    \caption{Physical implementation of the fault-tolerant \gls{pulp} cluster.}
    \label{fig:pie}
\end{figure}

\subsection{Experimental Setup}
Our evaluation setup consists of a \gls{pulp} cluster featuring 12 CV32E40P cores explained in \Cref{sec:pulp_cluster}, with an integrated \gls{hmr} unit as explained in \Cref{sec:integration}.
To evaluate the cluster's performance in different modes, the cluster was locked into the respective configuration before boot, avoiding unnecessary mode-switching overhead within the measurement. All performance evaluations were conducted in RTL simulation using QuestaSim, gathering performance data by using performance counters within the RTL or using simulation traces. We evaluated the split-lock and fault recovery performance by entering a \textit{mission-critical section} or \textit{performance section} twice to warm up the instruction caches and then collected the results corresponding to the hot cache measurement.
For implementation purposes, we target GlobalFoundries \SI{22}{\nano\meter} technology using Synopsys Design Compiler for synthesis (slow corner at $f_\mathrm{targ}=\SI{430}{\mega\hertz}$, $V_{DD}=\SI{0.72}{\volt}$, $T=\SI{-40}{\celsius},\SI{125}{\celsius}$) and Cadence Innovus for full-cluster Place \& Route in the same operating point.

\subsection{Physical Implementation}

\Cref{fig:layout} shows the post-layout implementation of our redundant \gls{pulp} cluster with all redundancy modes and \textit{rapid recovery} extension enabled, where we highlighted the cores with even and odd identifiers separately. \Cref{tab:area} shows the area occupation and the related overheads of all the available \gls{pulp} cluster configurations over the standard implementation, while \Cref{fig:pie_full} shows the cluster area breakdown. In the chosen corner, the \gls{pulp} cluster occupies \SI{0.660}{\milli\meter\squared}, where almost 27\% is the RISC-V cores, and around 29\% is the instruction cache. The \gls{hmr} unit, the wrapper in which we encapsulate all redundancy features, accounts for 9\% of the entire cluster area.

\Cref{fig:pie_hmr} shows the area breakdown of the \gls{hmr} unit, highlighting that the \textit{rapid recovery} extension accounts for 79\% of it. The \gls{tmr} voters and \gls{dmr} checkers account for 10\% and 5\% of the area occupation, respectively, while the shared control logic accounts for a 7\% area overhead. The \gls{hmr} unit is designed in a parametrized fashion, so it is possible to instantiate the unit without the hardware enhancement features for \textit{rapid recovery}, allowing a software-based recovery routine only. As shown in \Cref{tab:area}, the area overhead of the \gls{hmr} unit with only software recovery features is limited to almost $1\%$ of the \gls{pulp} cluster area, at a high cost in fault recovery performance. The fault-tolerant cluster could be synthesized at up to \SI{430}{\mega\hertz} operating frequency in the selected corner with no timing impact over the baseline implementation.

\begin{table}[t]
    \centering
    \caption{Summary of the performance provided by the proposed \gls{pulp} cluster in all the redundant available configurations.}
    \begin{tabular}{@{}llc rrrrr@{}}\toprule
                                        &  & & \multicolumn{1}{c}{base} & \multicolumn{1}{c}{DMR} & \multicolumn{1}{c}{TMR} & \multicolumn{1}{c}{DMR-R} & \multicolumn{1}{c@{}}{TMR-R} \\ \cmidrule{7-8}
                                        &  &  &  &  &  & \multicolumn{2}{c@{}}{Rapid Recovery} \\ \midrule
        \multicolumn{2}{@{}l}{MatMul Performance} & [MOPS @ \SI{430}{\mega\hertz}]            & 1165 & 617 & 414  & 617 & 414 \\
        \multicolumn{2}{@{}l}{\textit{SW-based MatMul Performance}} & \textit{[MOPS @ \SI{430}{\mega\hertz}]} & \textit{1165} & \textit{576} & \textit{351} & \multicolumn{1}{c}{-} & \multicolumn{1}{c@{}}{-} \\
        \multicolumn{2}{@{}l}{\acrshort{cfft} Performance} & [MOPS @ \SI{430}{\mega\hertz}] & 989 & 531 & 385 & 531 & 385 \\\midrule
        \multicolumn{2}{@{}l}{Recovery Latency} & [cycles]                  & \multicolumn{1}{c}{-} & \multicolumn{1}{c}{-} & 363 & 24 & 24 \\
        \multirow{3}{*}{Mission-Critical} & entry & \multirow{3}{*}{[cycles]} & \multicolumn{1}{c}{-} & 534 & 410 & 397 & 310 \\
                                          & exit main & & \multicolumn{1}{c}{-} & 22 & 23 & 22 & 23 \\
                                          & exit help & & \multicolumn{1}{c}{-} & 147 & 165 & 184 & 182 \\
        \multirow{2}{*}{Performance} & entry & \multirow{2}{*}{[cycles]} & \multicolumn{1}{c}{-} & 134 & 82 & 125 & 82 \\
                                          & exit & & \multicolumn{1}{c}{-} & 373 & 311 & 183 & 94 \\
        \bottomrule
    \end{tabular}
    \label{tab:results}
\end{table}

\subsection{Performance Evaluation}
\subsubsection{Compute Performance}
To evaluate the processing performance of the proposed fault-tolerant cluster, we initially target a highly parallel matrix-matrix multiplication benchmark. For evaluation, a size of 24$\times$24$\times$24 was chosen, meaning $2\times(24\times24\times24)=\SI{27648}{Ops}$, for optimal parallelization in all configurations. Furthermore, a highly parallel, quantized \gls{cfft} was evaluated, representing a typical workload for a \gls{scps}, e.g. used in radar processing. A length of 2048 was chosen, which amounts to $10\times \frac{n \log_2(n)}{2}=\SI{112640}{Ops}$.
\Cref{tab:results} summarizes the performance achieved by the fault-tolerant \gls{pulp} cluster, showing that at \SI{430}{MHz} our fault-tolerant cluster delivers a matrix-matrix multiplication compute performance of \SI{414}{MOPS} in \gls{tmode}, which increases to \SI{617}{MOPS} in \gls{dmode} and \SI{1165}{MOPS} in independent mode with all 12 cores available. Similarly, for the \gls{cfft}, the \gls{tmode} achieves a compute performance of \SI{385}{MOPS}, increasing to \SI{531}{MOPS} in \gls{dmode} and \SI{989} in independent mode. This performance boost of 1.8-1.9$\times$ for \gls{dmode} over the independent mode, and 2.5-2.8$\times$ for \gls{tmode} over independent mode justifies the choice to enable quick switching between different modes to balance the trade-off between performance and fault resilience. Furthermore, the hardware enhancements introduced for the \textit{rapid recovery} do not affect the compute performance of the \gls{pulp} cluster.

To compare our hardware-based solution, we implemented a software-based redundant execution of the same matrix-matrix multiplication kernel utilizing the parallel cores in the \gls{pulp} cluster for reliability. In the software-only approach, cores are similarly grouped (e.g., core 0 and core 6 in a \gls{dmr} approach), operating on the same chunk of the kernel data. At the end of the computation, the two cores check each other's results, raising an error in case of inconsistency. If the check results in no error, only one of the two cores (core 0, for example) validates the result storing the chunk in the memory. The use of software redundancy introduces significant overhead in designing the software, as it needs to be accounted for in each application manually. Further, it introduces a significant performance overhead due to memory contention (multiple cores accessing the same resources) and due to the mutual checking mechanism, resulting in 7\% performance overhead during \gls{dmr} execution and 11\% overhead in the \gls{tmr} case over the hardware lockstep approach. In addition, if the execution is corrupted in the \gls{dmr} software, the only way to recover from the fault is to repeat the computation, introducing additional non-negligible overhead that varies with the size of the executed kernel. Moreover, the overhead due to mutual checking in the software-only redundant execution depends on the size of the result.

\subsubsection{Recovery Performance}

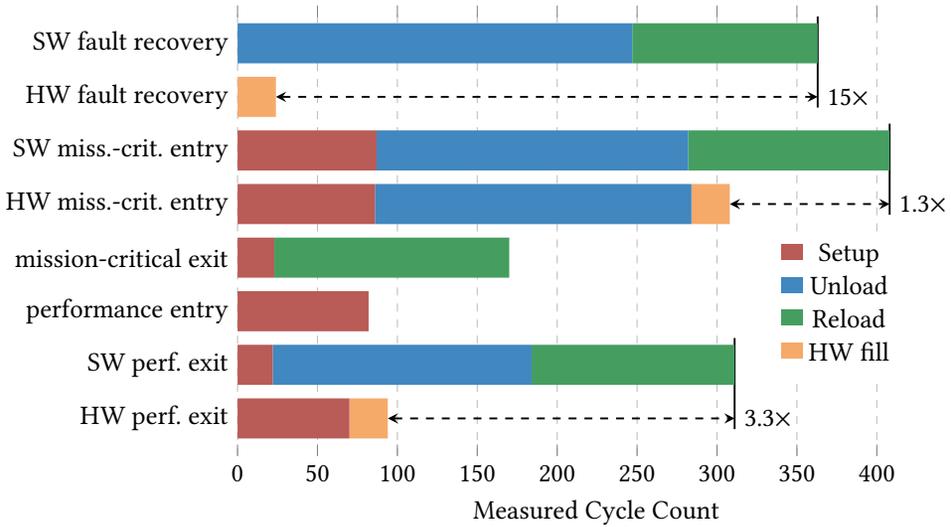
\begin{figure}[t]
    \centering
    \pgfplotstableread{ 
Label             Setup Unload Reload   {HW fill}
{HW perf. exit}                 70     0      0           24
{SW perf. exit}                 22  162   127           0
{performance entry}                 82     0      0           0
{mission-critical exit}        23      0    147           0
{HW miss.-crit. entry}    86    198      0          24
{SW miss.-crit. entry}    87    195    126           0
{HW fault recovery}   0      0      0          24
{SW fault recovery}   0    247    116           0
    }\testdata
    \begin{tikzpicture}
        \begin{axis}[
            xbar stacked,   
            xmin=0,         
            ytick=data,     
            legend style={at={(axis cs:335, 0.8)},anchor=south west, draw=none},
            yticklabels from table={\testdata}{Label},  
            bar width=15pt,
            xmajorgrids=true,
            grid style=dashed,
            ytick pos=left,
            width=0.8\columnwidth,
            height=7.5cm,
            xlabel={Measured Cycle Count},
            axis line style={draw=none}
        ]
            \addplot [fill=PULPred!80, draw opacity=0,] table [x=Setup, meta=Label,y expr=\coordindex] {\testdata};   
            \addplot [fill=PULPblue!80, draw opacity=0,] table [x=Unload, meta=Label,y expr=\coordindex] {\testdata};
            \addplot [fill=PULPgreen!80, draw opacity=0,] table [x=Reload, meta=Label,y expr=\coordindex] {\testdata};
            \addplot [fill=PULPorange!80, draw opacity=0,] table [x={HW fill}, meta=Label,y expr=\coordindex] {\testdata};
            \legend{Setup,Unload,Reload,{HW fill}}
            \draw[line width=0.25mm, black] [stealth-stealth, dashed] (24,6) -- (363,6) node [right] {15$\times$};
            \draw[line width=0.25mm, black] [] (363,7.5) -- (363,5.8);
            \draw [line width=0.25mm, black] [stealth-stealth, dashed] (308,4) -- (408,4) node [right] {1.3$\times$};
            \draw[line width=0.25mm, black] [] (408,5.5) -- (408,3.8);
            \draw [line width=0.25mm, black] [stealth-stealth, dashed] (94,0) -- (311,0) node [right] {3.3$\times$};
            \draw[line width=0.25mm, black] [] (311,1.5) -- (311,-0.2);
        \end{axis}
    \end{tikzpicture}
    \caption{Cycle count breakdown of the fault recovery and split-lock overheads.}
    \label{fig:bars_cycles}
\end{figure}

The software-based recovery routine, described in \Cref{sec:sw-resynch}, required 363 cycles to execute a recovery in \gls{tmode}, as shown in \Cref{tab:results}. Figure~\ref{fig:bars_cycles} highlights that this is split between the \textit{unload} section, requiring 247 cycles, after which it executes a synchronous clear of the core. Another 116 cycles are required for the \textit{reload} phase, reloading the cores' state from memory. The \textit{rapid recovery} hardware contributes positively by introducing a $15\times$ speed-up during the fault recovery, thus reducing it to just 24 cycles. This recovery time initially requires 4 cycles to set up the \textit{rapid recovery} controller, start the recovery routine, synchronously clear the cores, and send them the debug request. Additional 4 cycles are needed to receive the debug response from the cores, and 16 cycles to restore the \gls{pc}, \gls{rf}, and \glspl{csr} in parallel, after which they will continue processing. Furthermore, the \textit{rapid recovery} hardware also allows for fast fault recovery in \gls{dmode}, requiring the same fixed 24 cycles. The cycle-by-cycle backups of the cores' status make it possible to roll back the cores' execution to the closest safe state in time, restarting the execution from the last \gls{pc} that entered the execution stage.

\subsubsection{Split-lock Performance}
We evaluate the performance of the proposed split-lock mechanism by measuring the cycles required to enter or exit a \textit{mission-critical section} for the main and helper cores as well as a \textit{performance section}, as shown in \Cref{tab:results}. The entry measurement starts from the point the main core initiates the entry into the relevant section to the point the internal code is being executed.

For the \textit{mission-critical section}, \Cref{fig:bars_cycles} shows that the main core first sets up the registers for the intended mode within 87 cycles, during which time the helper core(s) can continue processing. At this point, the interrupt is issued to all cores, launching them into the \textit{unload} stage, where the cores save their state to memory and enter a barrier, requiring 198 cycles. Finally, during the \textit{reload} phase, the cores reload the main core's state from memory in the locked configuration, requiring 126 cycles, after which the cores execute the \textit{mission-critical section} code. When exiting the \textit{mission-critical section}, the main core goes directly into the subsequent software, requiring only 23 cycles to set the configuration and return. On the other hand, the helper cores need to execute the \textit{reload} procedure, requiring 147 cycles to continue the previously interrupted thread.

For the \textit{performance section}, \Cref{fig:bars_cycles} shows lower entry and exit cycle requirements than the \textit{mission-critical section}. Notably, entry into the \textit{performance section} only requires 82 cycles of setup. This is due to the cores only splitting apart and, if needed, updating a \gls{sp}, not fully reloading a known state as they can keep the current execution state. Furthermore, the \textit{performance section} exit drops the state of the assisting cores, relying fully on the main core's state, thus requiring significantly fewer cycles than the \textit{mission-critical section}.

Introducing the \textit{rapid recovery} hardware helps speed up entering a \textit{mission-critical section} as the software \textit{reload} stage can be replaced with a 24-cycle hardware fill from the continuously backed-up main core, meaning $\sim25\%$ less than the software-based approach for a $1.3\times$ speedup. Similarly, exiting a \textit{performance section} also makes use of the continually replicated main core state, leading to a $3.3\times$ speedup. Exiting a \textit{mission-critical section} or entering a \textit{performance section} behaves the same way as in the software-based approach.

\subsection{Reliability Assessment}
With the \gls{hmr} implementation outlined in this work, any error within a single core propagating to its outputs will be detected when mismatching with the paired cores, and corrective action will be taken. Furthermore, with the \textit{rapid recovery} extension enabled, core-internal signals backing up the \gls{rf} are also checked, reducing the likelihood of latent errors within the cores' states. While it aims to detect and correct all possible \glspl{set} happening at the cores' interface or within the cores themselves, the proposed redundancy approach does not tackle \glspl{seu} and single hard errors that might corrupt the data and instruction memory hierarchy, which we thus assume to be reliable.

To test the behavior when a fault occurs, a single bit of the cores' interface signals was inverted using a \texttt{force} command during the execution of a test within an RTL simulation. With the cores in the locked state, this confirms that errors were detected and that the designed corrective action was initiated. Furthermore, select state bits within the cores' \gls{rf} were flipped to confirm the proper behavior of the recovery implementations. All faults injected into the system were detected and corrected or overwritten before use, thus always leading to the correct termination of the running application.

The design was further subjected to a more extensive fault injection campaign, where all registers within a core were randomly subjected to faults. Running 7717 individual RTL simulations of the matrix multiplication performance benchmark in the \gls{tmode} configuration with software recovery, a single register was picked and flipped for each simulation run. All simulations terminated successfully, with 12\% of injected errors leading to a recovery, meaning the other faults were masked.

Other research has conducted extensive fault investigation of the CV32E40P core used in our system~\cite{asciolla_characterization_2020}, showing that many injected faults do not affect the program execution, with the most critical module being the controller module.
As only 12\% of injected errors lead to corrective action, this confirms the results of \citeauthor{asciolla_characterization_2020}~\cite{asciolla_characterization_2020}, where 44\%-100\% of injected errors lead to no correction, depending on the component.

\subsection{Recovery Use-case Analysis: Satellite Onboard Image Processing}
One of the most common use cases of multi-core \gls{scps} is satellite onboard image processing. Satellites' onboard sensors generate a large amount of data that heavily loads the data link and delays processing. When images captured onboard need to be processed through complex deep learning algorithms, the satellite first transmits images to the cloud computing data center of the ground station. Then, the data center operates on the received data using deep learning models and distributes the processing result to the user~\cite{wei_application_2019}. To reduce the overhead given by raw data transmission, the data processing can be directly moved onboard a spacecraft, only sending valuable information over the communication link. This requires the onboard processing systems to offer enough performance to process data while also offering reliability with quick recovery from incurring faults.

\begin{figure}[t]
    \centering
    \begin{tikzpicture}
        
        \begin{axis}[
            xmin=1000,
            xmax=1e8,
            ytick distance=0.2,
            ymin=0,
            ymax=1.25,
            samples=50,
            xmode=log,
            width=0.8\columnwidth,
            height=6.5cm,
            ylabel={Performance [GOPS]},
            xlabel={Fault Rate [\# faults/s]},
        ]
            \addplot[PULPpurple, ultra thick, dashed, domain=0.01:1e8] {27648/10203*0.43};
            \addplot[PULPred,  ultra thick, samples at={0.005138231,0.007707347,0.01156102,0.01734153,0.026012295,0.039018442,0.058527663,0.087791495,0.131687243,0.197530864,0.296296296,0.444444444,0.666666667,1,1.5,2.25,3.375,5.0625,7.59375,11.390625,17.0859375,25.62890625,38.44335938,57.66503906,86.49755859,129.7463379,194.6195068,291.9292603,437.8938904,656.8408356,985.2612534,1477.89188,2216.83782,3325.25673,4987.885095,7481.827643}] ({x/19266*430e6},{(27648/(19266+((x)*24))*0.43)});
            \addplot[PULPblue, ultra thick, samples at={0.005138231,0.007707347,0.01156102,0.01734153,0.026012295,0.039018442,0.058527663,0.087791495,0.131687243,0.197530864,0.296296296,0.444444444,0.666666667,1,1.5,2.25,3.375,5.0625,7.59375,11.390625,17.0859375,25.62890625,38.44335938,57.66503906,86.49755859,129.7463379,194.6195068,291.9292603,437.8938904,656.8408356,985.2612534,1477.89188,2216.83782,3325.25673,4987.885095,7481.827643}] ({x/19266*430e6},{(27648/(19266*(1+(x))))*0.43});
            \addplot[PULPorange,  ultra thick, samples at={0.005138231,0.007707347,0.01156102,0.01734153,0.026012295,0.039018442,0.058527663,0.087791495,0.131687243,0.197530864,0.296296296,0.444444444,0.666666667,1,1.5,2.25,3.375,5.0625,7.59375,11.390625,17.0859375,25.62890625,38.44335938,57.66503906,86.49755859,129.7463379,194.6195068,291.9292603,437.8938904,656.8408356,985.2612534,1477.89188,2216.83782,3325.25673,4987.885095,7481.827643}] ({x/28708*430e6},{(27648/(28708+((x)*24/2))*0.43)});
            \addplot[PULPgreen,  ultra thick, samples at={0.005138231,0.007707347,0.01156102,0.01734153,0.026012295,0.039018442,0.058527663,0.087791495,0.131687243,0.197530864,0.296296296,0.444444444,0.666666667,1,1.5,2.25,3.375,5.0625,7.59375,11.390625,17.0859375,25.62890625,38.44335938,57.66503906,86.49755859,129.7463379,194.6195068,291.9292603,437.8938904,656.8408356,985.2612534,1477.89188,2216.83782,3325.25673,4987.885095,7481.827643}] ({x/28708*430e6},{(27648/(28708+((x)*383/2))*0.43)});

            \legend{Baseline Performance, \gls{dcls} rapid, \gls{dcls}, \gls{tcls} rapid, \gls{tcls}}

            \draw [line width=0.25mm, cyan!70!black] [stealth-stealth, dashed] (1e6,0.03) -- (1e6,0.57) node [right] at (1.02e6,0.34) {> 6$\times$};
            \draw [line width=0.25mm, cyan!70!black] [stealth-stealth, dashed] (1e7,0.09) -- (1e7,0.31) node [right] at (1.02e7,0.2) {> 3$\times$};
        \end{axis}
    \end{tikzpicture}
    \caption{\gls{pulp} cluster performance degradation at increasing fault rate.}
    \label{fig:perfdegr}
\end{figure}
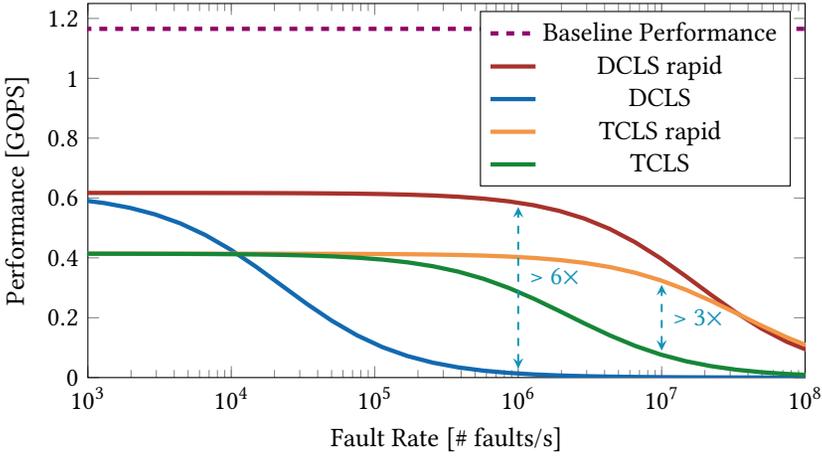

\Cref{fig:perfdegr} shows the performance degradation of our radiation-tolerant cluster when affected by multiple faults in all the available redundancy configurations. We considered a condition where the incurring faults happen during the execution time of a given application, with an increasing fault rate. In \gls{tmode}, if a fault affects one of the three grouped cores, the other two continue their operation. After some time, if another fault happens and affects another of the cores of the same group, it forces the entire group to perform a recovery. On the other hand, in \gls{dmode} without \textit{rapid recovery}, the cluster operation is restarted at each fault occurrence.

The analysis conducted shows that the worst-case scenarios are the software-based recovery procedures. In the \gls{dmode} case, a performance drop of $50\%$ happens with $\sim$\SI{2e4}{faults\per\second} injected. In the \gls{tmode}, the same performance drop happens with around \SI{2e6}{faults\per\second}. When using the \textit{rapid recovery} extension, the increased fault rate reduces the performance of the computing cluster much slower, as shown by the \gls{dcls} rapid and \gls{tcls} rapid curves in \Cref{fig:perfdegr}. The \textit{rapid recovery} feature allows for more than $6\times$ performance speedup during fault recovery over the software-based procedure at \SI{e6}{faults\per\second}, with the performance of the \gls{dcls} rapid case remaining almost constant. Similarly, \gls{tcls} rapid delivers $3\times$ speedup over the software-based recovery with \SI{e7}{faults\per\second} injected. Interestingly, the \gls{dcls} rapid performs better than \gls{tcls} rapid until the fault rate reaches around \SI{3e7}{faults\per\second}. As the fault rate increases further, the performance in \gls{tcls} rapid is the best because, with \gls{tcls}, we do not necessarily need to recover immediately from a fault. If one of the three grouped cores is faulty, the computation can continue with the other two cores, and a re-synchronization is only needed if another fault affects the same group of cores. In the \gls{dcls} rapid case, a re-synchronization is needed as soon as a fault happens. With \textit{rapid recovery} features, we could observe a $50\%$ performance degradation with a fault rate of around \SI{2e7}{faults\per\second} and \SI{4e7}{faults\per\second} for \gls{dcls} rapid and \gls{tcls} rapid, respectively. Furthermore, the conducted analysis shows that the proposed fault-tolerant \gls{pulp} cluster allows for higher computing performance depending on the criticality of the application.

In independent mode, we only have two ways to determine that a fault happened. The first case is instructions faults, meaning incoming faults compromise the executed instructions sending the victim core in an unknown state. In this case, the faulty core stalls, and an external watchdog restarts the cluster operation. The second scenario is that of data faults. Suppose a fault affects the calculation of the cluster cores. In that case, we can only identify that an error happens if the host core knows the computation results a priori and can verify the results produced by the cluster computation. Both these two scenarios imply a significant overhead, similar to or worse than the \gls{dcls} software-based recovery.

\begin{figure}[t]
    \centering
\begin{tikzpicture}
    \pgfmathsetmacro{\numEx}{1}
    \pgfmathsetmacro{\numExDiv}{1}
    \pgfmathsetmacro{\xmin}{-3.2}
    \pgfmathsetmacro{\xmax}{0.2}
    \pgfmathsetmacro{\ymin}{-4}
    \pgfmathsetmacro{\ymax}{3}
    \pgfmathsetmacro{\zmin}{-10.5}
    \pgfmathsetmacro{\zmax}{0}
    \pgfmathsetmacro{\ensingle}{1/1000}
    \pgfmathsetmacro{\scalesize}{1.0}
    \begin{axis}[
        name=dmr,
        unit vector ratio*=\scalesize 1 1,
        xmin=\xmin,
        xmax=\xmax,
        ymin=\ymin,
        ymax=\ymax,
        xtick={-6, -5, -4, -3, -2, -1, 0},
        xticklabels={$10^{-6}$, $10^{-5}$, $10^{-4}$, $10^{-3}$, $10^{-2}$, $10^{-1}$, $10^0$},
        ytick={-4, -2, 0, 2, 4},
        yticklabels={, $10^{-2}$, $10^0$, $10^2$, $10^4$},
        minor y tick num=1,
        ylabel={Execution Time [\si{\second}]},
        title=\gls{dcls},
        title style = {text depth=0.5ex},
        small,
        view={0}{90},
     ]
        \addplot3 [surf,
            shader=interp,
            samples=32,
            point meta min=\zmin,
            point meta max=\zmax,
        ] {ln(max(\ensingle,10^x*10^y)*(10^y/\numEx/2)/\numExDiv)/ln(10)};
    \end{axis}
    \begin{axis}[
        name=dmrr,
        at={($ (0.5cm,0cm) + (dmr.south east) $)},
        unit vector ratio*=\scalesize 1 1,
        xmin=\xmin,
        xmax=\xmax,
        ymin=\ymin,
        ymax=\ymax,
        xtick={-6, -5, -4, -3, -2, -1, 0},
        xticklabels={$10^{-6}$, $10^{-5}$, $10^{-4}$, $10^{-3}$, $10^{-2}$, $10^{-1}$, $10^0$},
        ytick={-4, -2, 0, 2, 4},
        yticklabels=\empty,
        minor y tick num=1,
        title=\gls{dcls} rapid,
        title style = {text depth=0.5ex},
        small,
        view={0}{90},
     ]
        \addplot3 [surf,
            shader=interp,
            samples=32,
            point meta min=\zmin,
            point meta max=\zmax,
        ] {ln((56e-9)*max(\ensingle,10^x*10^y)/\numExDiv))/ln(10)};
    \end{axis}

    \begin{axis}[
        name=tmr,
        at={($ (0.5cm,0cm) + (dmrr.south east) $)},
        unit vector ratio*=\scalesize 1 1,
        xmin=\xmin,
        xmax=\xmax,
        ymin=\ymin,
        ymax=\ymax,
        xtick={-6, -5, -4, -3, -2, -1, 0},
        xticklabels={$10^{-6}$, $10^{-5}$, $10^{-4}$, $10^{-3}$, $10^{-2}$, $10^{-1}$, $10^0$},
        ytick={-4, -2, 0, 2, 4},
        yticklabels=\empty,
        minor y tick num=1,
        xlabel={Error rate [faults/\si{\second}]},
        title=\gls{tcls},
        title style = {text depth=0.5ex},
        small,
        view={0}{90},
        ]
        \addplot3 [surf,
            shader=interp,
            samples=32,
            point meta min=\zmin,
            point meta max=\zmax,
        ] {ln((890e-9)*max(\ensingle,(10^x/2)*10^y)/\numExDiv)/ln(10)};
    \end{axis}

    \begin{axis}[
        name=tmrr,
        at={($ (0.5cm,0cm) + (tmr.south east) $)},
        unit vector ratio*=\scalesize 1 1,
        xmin=\xmin,
        xmax=\xmax,
        ymin=\ymin,
        ymax=\ymax,
        xtick={-6, -5, -4, -3, -2, -1, 0},
        xticklabels={$10^{-6}$, $10^{-5}$, $10^{-4}$, $10^{-3}$, $10^{-2}$, $10^{-1}$, $10^0$},
        ytick={-4, -2, 0, 2, 4},
        yticklabels=\empty,
        minor y tick num=1,
        title=\gls{tcls} rapid,
        title style = {text depth=0.5ex},
        small,view={0}{90},colorbar,
        colorbar style={
            ylabel={Runtime Overhead},
            ytick={-10, -9, -8, -7, -6, -5, -4, -3, -2, -1, -0.2},
            yticklabels={$<100$ns, 100ns, 1us, 10us, 100us, 1ms, 10ms, 100ms, 1s, 10s, $>100$s},
            yticklabel style={
                /pgf/number format/.cd,
                fixed,
                fixed zerofill,
                precision=1,
            }
        },
    ]
        \addplot3 [surf,
            shader=interp,
            samples=32,
            point meta min=\zmin,
            point meta max=\zmax,
        ] {ln((56e-9)*max(\ensingle,(10^x/2)*10^y)/\numExDiv)/ln(10)};
    \end{axis}

\end{tikzpicture}
    \caption{\gls{pulp} cluster recovery cost from incurring faults in the available redundant modes.}
    \label{fig:rtoverhead}
\end{figure}
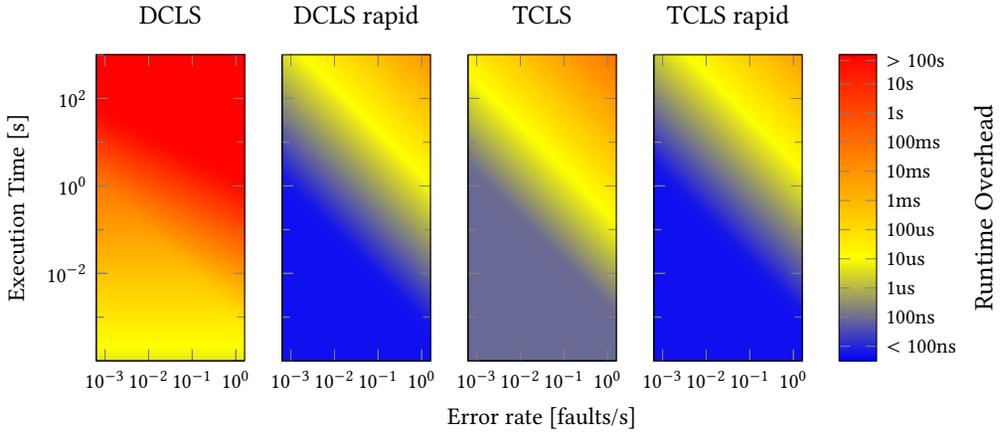

The analysis conducted so far helps understand the behavior of the system when exposed to varying fault rates. The fault rate corresponding to the worst-case estimation for a low-earth orbit device exposure to cosmic rays is in the order of 1.1-\SI{1.7e-3}{faults\per\second} (meaning 100/150 faults/day) and can vary with the flying orbit of the system and with the technology node~\cite{engel_predicting_2006, di_mascio_open-source_2021}. Furthermore, the probability that faults happening in sequence affect the same group of cores is low. In this scenario, the probability that the chip is corrupted by only one single fault during the entire execution of an application is high, even if the application runs for several seconds. If the fault hits one of three cores configured in \gls{tcls} with or without \textit{rapid recovery}, this fault may not affect the execution time of the application. Here, the two remaining cores can continue their operation without interruption until another fault hits the same group.

On the other hand, deeper considerations are needed if the system is configured in \gls{dcls}. With \textit{rapid recovery}, a single fault recovery has no impact on an application that runs for more than thousands of clock cycles (meaning at least \SI{2}{\micro\second} at the considered frequency). The motivation is that the \textit{rapid recovery} extension guarantees 24 clock cycles (approximately \SI{56}{\nano\second} at \SI{430}{\mega\hertz}) to complete the entire recovery procedure. Thus, if an application lasts several seconds, the \gls{dcls} rapid configuration does not introduce any significant degradation in the execution time.

The same does not hold in the case of \gls{dcls}. In this configuration, the only way to recover from a fault hitting a group of cores is to restart the application execution. The consequence is that the time to recovery varies depending on two factors: the moment the fault occurs during execution and the duration of the application. The moment the fault corrupts the executed application (if at the beginning, in the middle, or at the end) impacts the time to recovery significantly. In the worst case, the fault happens close to the end of the execution. In this situation, the time to recovery equals the time of the application itself because it must be repeated almost entirely. The second factor, the execution time of the application (hundreds of microseconds or tens of seconds), further affects the time to recovery. A critical operation may have a strict deadline that is only slightly longer than the required computation time. In this case, the unpredictability of the recovery time can severely impact the real-time capabilities of a system, potentially leading to a failure of the mission. \rebuttal{The ISO 26262 standard provides a definition for the Fault Tolerant Time Interval (FTTI), which is the duration required to identify a fault within the system, initiate a response to the fault, and return the system to a safe state prior to the occurrence of a hazardous event. FTTI is comprised of three components: the time needed for fault detection (Diagnostic Test Interval), the time taken to respond to the fault (Fault Reaction Time), and the time required to re-establish a safe operational state (Safe Tolerance Time) \cite{watzenig_functional_2017}. The precise definition of these parameters is contingent on the specific application executed by the system and should be minimized to ensure the system operates safely across various scenarios.}

In \Cref{fig:rtoverhead}, we show how the recovery cost from incurring errors affects the execution of an application when the error rate ranges between \SI{e-3}{faults\per\second} and \SI{1}{faults\per\second}. We show the recovery cost through a runtime overhead expressed in seconds, highlighted on the color bar on the right, and such overhead is computed taking into account a minimum of \SI{e-3}{faults} happening per application execution, on average.

For the \gls{dcls} configuration, we have considered that the incurring faults happen in the middle of the execution, which is the fault occurrence expected value. The consequence is that for every fault affecting the system, the overhead introduced for the recovery is half the execution time. Under these assumptions, the plots show that the \gls{pulp} cluster configured in \gls{dcls} without the \textit{rapid recovery} has a minimum runtime overhead of \SI{10}{\micro\second}. In particular, at a \SI{1e-3}{faults\per\second} rate, if the executed application requires 1-\SI{100}{\second}, the runtime overhead equals the execution time of the given application, hence being not suitable for real-time executions. Conversely, when configured in \gls{dcls} rapid, \gls{tcls}, and \gls{tcls} rapid, there is a wide range where the recovery cost is negligible. The result is that the runtime overhead due to the recovery is always lower than \SI{10}{\milli\second}, also at higher fault rates. \rebuttal{It is crucial to highlight that the rapid recovery feature enables the DCLS configuration to align with real-time constraints. This allows for a balance between computational performance and the capacity for real-time fault recovery, facilitating the reduction of the FTTI.} It is important to note that the \textit{rapid recovery} feature makes it possible for the \gls{dcls} configuration to meet real-time constraints, allowing for a trade-off between computing performance and real-time fault recovery capabilities.

%% file: src/sections/05_Discussion.tex
\section{State of the Art Comparison}\label{section:Discussion}

\input{src/tables/soa_table}

\Cref{tab:soa_table} shows the details of our implementation alongside the State of the Art. Compared with ARM, their TCLS~\cite{iturbe_arm_2019} implementation features a split-lock mechanism that is based on resetting the system to configure it in independent or \gls{tmode}. Furthermore, the recovery routine they propose takes 2351 clock periods to conclude, meaning $6.5\times$ slower than our software-based \gls{tmode}. ARM's DCLS~\cite{iturbe_soft_2016} also features split-lock functionalities decided during system reset. Our split-lock allows for higher flexibility and performance, making the cluster capable of runtime switching between independent and \gls{dmode} or \gls{tmode} in 681 and 597 clock cycles, respectively, with further $1.17\times$ and $1.08\times$ reduction with the \textit{rapid recovery} hardware support. Moreover, ARM's TCLS implementation introduces $27\%$ area overhead over the single core implementation, while the \gls{hmr} unit with \textit{rapid recovery} features proposed in this work introduces just $9\%$ area overhead over the baseline implementation. 

\citeauthor{kempf_adaptive_2021}~\cite{kempf_adaptive_2021} propose a dual-core adaptive runtime-selectable lockstep processor based on a LEON core with internal modifications for instruction comparison. The fault recovery is software-based, with a time-to-recovery that is application dependent. Our \textit{rapid recovery} hardware extension on the other hand allows for fixed 24 cycles to recover from occurring faults. Also, the area overhead of the solution proposed by~\citeauthor{kempf_adaptive_2021} accounts for $21.3\%$ in terms of CLBs over the regular LEON implementation, while our full \gls{hmr} unit accounts for just $9\%$ of area overhead over a standard 12-cores \gls{pulp} cluster, with a similar dual-core configuration needing only $8.4\%$.

CEVERO~\cite{silva_cevero_2020} proposes a DCLS system with no split-lock capabilities and with a similar \textit{rapid recovery} hardware extension to ours, with a recovery procedure that takes $1.67\times$ longer than ours. Furthermore, CEVERO disregards copying the \glspl{csr}, which are fundamental in defining the complete recovery state of the core, and the backup copies of \gls{pc} and \gls{rf} are not protected by error correction, meaning there is no guarantee that the backup state is reliable.

\citeauthor{shukla_low-overhead_2022}~\cite{shukla_low-overhead_2022} present a quad-core RISC-V-based processor re-configurable for DCLS operation, introducing up to $17.9\%$ area overhead over the base implementation, while our \gls{hmr} unit offers more flexibility by introducing just $9\%$ area overhead over a standard 12-cores \gls{pulp} cluster. In addition, \citeauthor{shukla_low-overhead_2022} rely on saving just the last executed \gls{pc} for the recovery, which is insufficient to determine the entire state of the grouped cores.

In \Cref{tab:soa_table}, we also compare our design with other works, such as SHATKI-F~\cite{gupta_shakti-f_2015}, Duck Core~\cite{li_duckcore_2021}, and \citeauthor{gkiokas_fault-tolerant_2019}~\cite{gkiokas_fault-tolerant_2019}. The first two works propose modifications to RISC-V cores' internal microarchitecture for pipeline rollback. On the other hand, the third proposes extending a RISC architecture with triple repetition of all the pipeline stages until the execution stage. The results produced by the three execution stages are then voted and propagated to the memory and write-back stages of the pipeline. The latter two pipeline stages do not enforce any redundancy technique to guarantee consistency, thus leaving the memory and write-back stages unprotected. The proposed solutions show a valuable approach that leads to just 3-to-0 clock cycles to perform a fault recovery, allowing for a single core to be reliable without the need for redundant grouping, thus saving resources. In contrast, the extensive modifications required by the internal architecture of the core can significantly change the behaviour, compromising its formal verification.

We also compared our work with SafeDE~\cite{bas_safede_2021} and SafeDM~\cite{bas_safedm_2022}, proposing hardware monitors to enforce diverse redundancy in a multicore RISC-V system, tackling the coverage of common cause failures in redundant threads. In SafeDE, the only way to understand if the diversity exists in the executed thread is to compute the distance, in terms of instructions count, between the head and the trailer core, with the trailer core being stalled for the required clock cycles to respect the diversity constraint. However, there is no description of how to mitigate the case in which one of the two executions is corrupted by a fault, so we assume that the two cores are reset and forced to re-execute the code. Similar considerations apply to SafeDM, where it is unclear if a fault between the two redundant executions can even be detected.

Finally, \Cref{tab:soa_table} also shows qualitative comparisons to the industry-standard approach of \gls{radhard} technology augmented with \gls{ecc} for memories. This approach, adopted by Gaisler~\cite{hijorth_gr740_2015}, BAE~\cite{berger_quad-core_2015}, Ramon Chips~\cite{ginosar_rc64_2016}, and the DAHLIA project with the NG-ultra~\cite{danard_ng-ultra_2022}, significantly simplifies integration of reliability. However, these devices generally feature lower performance and lower efficiency than counterparts built on commercial technologies and often come at a significant price premium due to the use of \gls{radhard} technology. Furthermore, in contrast to our solution, they offer no way to disable these reliability methods in case they are not needed, sacrificing performance in case they are still used in these cases. If a large-scale device deployment requires certain nodes with reliability and certain nodes without, either different devices need to be used, requiring additional development overhead, or certain nodes pay in performance, efficiency, and cost. Finally, architectural solutions can more easily benefit from advances in technology, as these can more easily be ported to the most modern commercial technologies, not requiring additional development of new radiation-hardened libraries and design kits.

%% file: src/tables/soa_table.tex
\begin{table}[t]
    \centering
    \caption{State of the Art comparison table for qualitative comparison on the top side, and quantitative comparison on the bottom side.}
        \begin{adjustbox}{width=1\textwidth,center=\textwidth}
        \begin{threeparttable}
        \begin{tabular}{@{}l ccccrrrrc@{}}\toprule
             & \multirow{2}{*}{\gls{isa}} & Reliability & Fault & Split- & \multicolumn{1}{c}{Area} & \multicolumn{1}{c}{Recovery} & \multicolumn{1}{c}{Frequency} & \multicolumn{1}{c}{Area} & Open-
            \\ 
            & & Method & Recovery & Lock & \multicolumn{1}{c}{Increase} & \multicolumn{1}{c}{Cycles} & \multicolumn{1}{c}{[\si{\mega\hertz}]} & \multicolumn{1}{c}{[\si{\milli\meter\squared}]} & Source \\\midrule
            \multirow{4}{*}{\textbf{This Work}} & \multirow{4}{*}{\textbf{RISC-V}} & \multirow{2}{*}{\textbf{\gls{dmode}}} & \textbf{SW} & \multirow{4}{*}{\cmark} & \textbf{1.4\%} & \textbf{-} & \multirow{4}{*}{\textbf{430}} & \textbf{0.608} & \multirow{4}{*}{\cmark} \\
            &  &  & \textbf{HW} &  & \textbf{7.8\%} & \textbf{24} &  & \textbf{0.657} &  \\
            &  & \multirow{2}{*}{\textbf{\gls{tmode}}} & \textbf{SW} &  & \textbf{1.8\%} & \textbf{363} &  & \textbf{0.605} & \\
            &  &  & \textbf{HW} &  & \textbf{8.3\%} & \textbf{24} &  & \textbf{0.654} & \\
            ARM DCLS~\cite{iturbe_soft_2016} & ARM & \gls{dcls} & SW & \cmark \tnote{1} & - & App. dep. & - & - & \xmark \\
            ARM TCLS~\cite{iturbe_arm_2019} & ARM & \gls{tcls} & SW & \cmark \tnote{1} & 27\% & 2351 & 314 & 3.640 & \xmark \\
            \citeauthor{shukla_low-overhead_2022}~\cite{shukla_low-overhead_2022} & RISC-V & \gls{dcls} & HW & \cmark \tnote{1} & 17.9\% & 1 & 41 & 0.223 & \xmark \\
            \citeauthor{kempf_adaptive_2021}~\cite{kempf_adaptive_2021} & SPARC & \gls{dcls} & SW & \cmark & 21.3\%\tnote{ 2} & - & 100 & - & \xmark \\
            CEVERO~\cite{silva_cevero_2020} & RISC-V & \gls{dcls} & HW & \xmark & - & 40 & - & - & \cmark \\
            SHAKTI-F~\cite{gupta_shakti-f_2015} & RISC-V & block-level \gls{dmr} & HW & \xmark & 20.5\% & 3 & 330 & 0.325 & \xmark \\
            \citeauthor{gkiokas_fault-tolerant_2019}~\cite{gkiokas_fault-tolerant_2019} & RISC & block-level \gls{tmr} & HW & \xmark & 15\%\tnote{ 2} & 0 & 50.56 & - & \xmark \\
            Duck Core~\cite{li_duckcore_2021} & RISC-V & \gls{ecc} & HW & \xmark & 0.7\%\tnote{ 2} & 3 & 50 & - & \xmark \\
            SafeDE~\cite{bas_safede_2021} & RISC-V & diverse redundancy & HW & \xmark & 0.79\%\tnote{ 2} & - & - & - & \cmark \\
            SafeDM~\cite{bas_safedm_2022} & RISC-V & diverse redundancy & HW & \xmark & 3.4\%\tnote{ 2} & - & - & - & \cmark \\
            \bottomrule
        \end{tabular}
        \begin{tablenotes}
            \item[1] Mode switching only possible during reset
            \item[2] Overhead in CLBs or LUTs
        \end{tablenotes}
        \end{threeparttable}
        \end{adjustbox}
    
    \label{tab:soa_table}
\end{table}

%% file: src/sections/06_Conclusion.tex
\section{Conclusion}\label{section:Conclusion}
In this work, we enhanced a \gls{pulp} cluster with fault-tolerant capabilities introducing a novel Hybrid Modular Redundancy approach. Our solution allows for flexible on-demand dual-core and triple-core lockstep grouping with dynamic runtime split-lock capabilities and hardware/software-based recovery approaches, being the first system to integrate similar functionalities on an open-source RISC-V-based compute device for finely-tunable reliability vs. performance
trade-offs. The hardware-based recovery provides rapid fault recovery in just 24 clock cycles with just $\sim9\%$ area overhead over the baseline implementation, while the software-based recovery takes 363 clock cycles introducing just $1.3\%$ area overhead. The proposed runtime split-lock mechanism allows for entering and exiting one of the available redundant modes with minimal performance loss, allowing execution of mission-critical portions of code or sections requiring performance while otherwise in reliable mode with <300 and <200 cycles of configuration overhead, respectively.